\documentclass[aps]{revtex4}
\usepackage{epsfig}
\usepackage{graphicx}
\usepackage{amsmath}
\begin{document}                
%
%
\title{
Entanglement-Assisted Coherent Control in
Nonreactive Diatom-Diatom Scattering}
\author{Jiangbin Gong, Moshe Shapiro \footnote{Permanent address:
Chemical Physics Department, The Weizmann Institute of Science, Rehovot, 
Israel 76100}, and Paul Brumer}
\affiliation{Chemical Physics Theory Group,\\ Department of Chemistry,\\
University of Toronto\\ 
Toronto, Canada  M5S 3H6}
\date{\today}

\begin{abstract}

Intriguing quantum effects that
result from entangled molecular rovibrational states are shown to
provide a novel means for controlling
both differential and total collision cross sections in identical
particle diatom-diatom scattering. Computational results
on elastic and inelastic scattering of para--H$_2$
and para--H$_2$ are presented, with the collision energy ranging from $400$ cm$^{-1}$
to the ultracold regime. The experimental realization and
 possible extension to other systems are discussed.
\end{abstract}

\maketitle

\section{Introduction}

Recent years have witnessed an increasing interest in controlling
molecular and atomic processes \cite{paulreview,rice,rabitz}. The central
principle of Coherent Control (CC) teaches that such control can be
achieved by using coherent laser fields to induce constructive or
destructive interference  between multiple indistinguishable pathways to
the same final state.  The yield of desired or undesired product can thus
be enhanced or suppressed by manipulating the resultant quantum
interference terms. Thus far,  most experimental and theoretical studies
of CC of atomic and molecular processes have focused on unimolecular
reactions or half collisions,  such as photodissociation, photoionization
and photoassociation.

By contrast,  controlling full collision processes such as atom-atom
scattering,  atom-molecule scattering, and molecule-molecule scattering
remains a significant challenge. Existent studies along this direction
have largely employed a number of approaches
\cite{sanchez,fedichev,dcfield,paulsca,lasercat}. The first
\cite{sanchez,fedichev,dcfield} actively manipulates coherent light
sources to alter the interaction potential of the colliding particles.
For example, in ultracold atom-atom collisions, the scattering cross
sections can be extensively modified by using laser fields that are in
near resonance with one of the excited molecular electronic states
\cite{sanchez,fedichev}, or by employing a d.c. field that induces an
additional dipole-dipole interaction potential \cite{dcfield}. However, it
appears difficult to use this approach to control atom-molecule or
molecule-molecule scattering,  since the number of degrees of freedom is
much larger than that in atom-atom scattering. A second approach is laser
catalysis \cite{lasercat} in which the laser field photons serve as a
catalyst. In this case the laser accelerates (or suppresses) 
the reaction by inducing a barrier hopping process, via 
a virtual excitation to a bound excited state, 
while remaining unchanged at the end of the event. The third  approach,
which is the subject of this study,  is coherent control \cite{paulsca}.
Although the general theory of collisional CC has been worked out
\cite{paulsca} and conditions for control established \cite{alex} the
experimental problem remains as to how one can introduce
the required adjustable
quantum phases and amplitudes into coherent scattering processes in a
fashion so as to introduce control based upon quantum interference.
That is, several recent studies on CC of collisions in model
scattering systems \cite{alex}  make clear that, for collisions of
non-identical molecules, coherent control requires that one establish
subtle quantum correlations between the translational and internal degrees
of freedom of the reactants. Setting up the required initial states
presents an experimental challenge,  since it requires coherent
matter-wave beams in which the internal states of the molecule are
correlated in a precise way with the translational states.

Such correlated translational-internal states are not
required, however, if one attempts coherent control in the collision of identical
molecules, i.e., AB + AB. Rather,  it suffices \cite{pauljcp} to prepare the
reactants in superpositions of initial rovibrational states. In this case,
in accord with Ref. \cite{pauljcp}, the collision cross sections can be
controlled by manipulating the character of the initial coherent
superposition states in which the molecules are prepared.

In this paper we examine the coherent control of identical particle
scattering in detail, with para--H$_2$ + para--H$_2$ as an example. In
doing so we approach this issue from a different perspective from that in
our earlier work \cite{pauljcp}, demonstrating deep connections to the
construction and use of {\em entangled states}, a topic of great recent
interest in quantum information science \cite{qcom}. That is, we show that
the scattering amplitudes, and the resultant differential and total
collision cross sections, depend strongly on quantum entanglement embedded
in the initial rovibrational  state. In essence, entanglement-induced quantum
effects are shown to occur without the explicit preparation of entangled states.
This provides a novel means of controlling both the the differential and
total collision cross sections in identical diatom-diatom scattering.

This paper is organized as follows.  In Sec. \ref{s5-2}, we introduce some
concepts related to quantum entanglement between identical particles and
then address the theory of identical diatom-diatom scattering with
particular reference to entangled molecular rovibrational states as the
incoming asymptotic states. The theoretical formalism is presented 
in Sec. \ref{formalism} for
para--H$_2$ + para--H$_2$ system,  without loss of much generality.
Results for the scattering of this system are shown in Sec. \ref{s5-3}
where detailed computational results on the role of entangled
rovibrational states in both elastic and inelastic scattering, with the
collision energy ranging from $400$ cm$^{-1}$ to the ultracold regime, are
explored. A discussion is provided in Sec. \ref{s5-4}.

\section{AB + AB Scattering}
\label{s5-2}

\subsection{Identical Particle Entanglement}

Quantum entanglement is a striking feature of quantum physics.
Qualitatively, two particles are said to be quantum entangled if their
total wavefunction is inseparable.
In particular, if two particles are entangled
then neither of them possesses a complete set of properties, since
measuring one of them would collapse the total wavefunction and thus
affect the properties of the other. Quantum entanglement is a fundamental
issue of considerable theoretical interest, and plays a key role in
various modern areas such as quantum teleportation \cite{qtele,kurizki},
quantum cryptography, and quantum computing \cite{qcom}.

In the case of identical particles the total wavefunction, including all
degrees of freedom, is always symmetrized or anti-symmetrized with respect
to particle exchange. Hence, even if the total wavefunction of two
identical particles is simply obtained by symmetrizing (or
anti-symmetrizing) a separable state, it appears to be unfactorizable, and
hence apparently entangled. This is, in fact, misleading and hints that
quantum entanglement between identical particles is somewhat more subtle
than quantum entanglement between non-identical particles. In a recent
discussion, Ghirardi et al \cite{ghirardi}  have given clear definitions of
quantum entanglement between two identical particles: (i) two identical
fermions are entangled if  the total wavefunction cannot be obtained by
anti-symmetrizing a factorized state, and (ii) two identical bosons are
entangled if the total wavefunction cannot be obtained by symmetrizing a
factorized product of two orthogonal states and if the total wavefunction
is not a product of the same state for the two particles. Under these
circumstances, the state is entangled and each of the particle pairs does
not possess a full set of properties.

One intriguing aspect of quantum entanglement between identical particles,
relevant to the discussion below, is that it can force different
permutation symmetries on some degrees of freedom. For example, consider a
system of two freely moving identical spin $1/2$ particles, whose spin up
and spin down states are represented by $|\uparrow_{1}\rangle$,
$|\downarrow_{1}\rangle$,
 $|\uparrow_{2}\rangle$, $|\downarrow_{2}\rangle$. If the total spin is
zero, then the two spins are entangled and the resulting (``EPR state")
$|\psi\rangle$ is given by
\begin{eqnarray}
|\psi\rangle=\frac{1}{\sqrt{2}}(|\uparrow_{1}\rangle\otimes|\downarrow_{2}\rangle
-|\downarrow_{1}\rangle\otimes|\uparrow_{2}\rangle).
\label{spinEPR}
\end{eqnarray}
Note that $|\psi\rangle$ in Eq. (\ref{spinEPR}) cannot be obtained by
anti-symmetrizing a factorizable {\em total} wavefunction (since such an
anti-symmetrized state should also describe the translational motion) of
two spin $1/2$ particles. Clearly, $|\psi\rangle$ in Eq. (\ref{spinEPR})
acquires a factor of $-1$  upon permutation. Since the permutation
symmetry for the total wavefunction including all degrees of freedom is
$-1$ for identical fermions, the permutation symmetry of the spatial
degrees of freedom must be $+1$.  Thus, in a simple bound system such as
para--H$_2$, the rotational quantum number has to be even.   Likewise,
in unbound cases such as electron-electron or proton-proton scattering,
the differential cross sections do depend strongly on how the two spins
are entangled, even if the scattering potential is spin independent, e.g.,
a pure Coulomb potential \cite{taylor51}. This selection of specific
permutation symmetry combinations due to entanglement of a subset of the
system degrees of freedom is central to the results below.

\subsection{The Role of Entangled Rovibrational States in Identical Particle
Scattering}

Consider identical particle scattering. We focus on the diatom-diatom case
AB + AB, but the considerations are general. If $|j,m,v\rangle$ denotes an
eigenstate of the AB diatom with angular momentum quantum number $j$,
angular momentum projection quantum number $m$ and vibrational quantum number
$v$, then a typical entangled ro-vibrational state of AB + AB is of the form
\begin{eqnarray}
\langle {\bf r}_{1}, {\bf r}_{2}|\psi(j_{1}m_{1}v_{1}j_{2}m_{2}v_{2})\rangle_{\alpha, \beta}& \equiv&
\cos(\alpha) \langle {\bf r}_{1} |j_{1}, m_{1}, v_{1}\rangle  \langle
{\bf
r}_{2}
|j_{2}, m_{2}, v_{2}\rangle \nonumber \\ & & +
\sin(\alpha) \exp(i\beta) \langle {\bf r}_{1}|j_{2}, m_{2}, v_{2}\rangle
\langle {\bf r}_{2}|j_{1}, m_{1}, v_{1}\rangle.
\label{entangled}
\end{eqnarray}
Of specific interest later below are the entangled states
\begin{eqnarray}
|\psi(j_{1}m_{1}v_{1}j_{2}m_{2}v_{2})\rangle_{\pm}&\equiv&
\frac{1}{\sqrt{2}}(\langle {\bf r}_{1} |j_{1}, m_{1}, v_{1}\rangle  \langle {\bf r}_{2}
|j_{2}, m_{2}, v_{2}\rangle \nonumber \\
&& \pm \langle {\bf r}_{1}|j_{2}, m_{2}, v_{2}\rangle
\langle {\bf r}_{2}
|j_{1}, m_{1}, v_{1}\rangle),
\label{definepm}
\end{eqnarray}
where the only
distinction between the two states $|\psi(j_{1}m_{1}v_{1}j_{2}m_{2}v_{2})\rangle_{\pm}$
is the relative phase $\beta$ ($0$ or $\pi$) between
the two participating states $\langle {\bf r}_{1}|j_{1}m_{1}v_{1}\rangle\langle{\bf
r}_{2}|j_{2}m_{2}v_{2}\rangle
$ and $\langle{\bf r}_{1}|j_{2}m_{2}v_{2}\rangle\langle {\bf r}_{2}|j_{1}m_{1}v_{1}\rangle$.
This difference is significant insofar as
$|\psi(j_{1}m_{1}v_{1}j_{2}m_{2}v_{2})\rangle_{+}$
is invariant upon permutation of ${\bf r}_{1}$ and ${\bf r}_{2}$, whereas
$|\psi(j_{1}m_{1}v_{1}j_{2}m_{2}v_{2})\rangle_{-}$ acquires a factor of $-1$
upon permutation of ${\bf r}_{1}$ and ${\bf r}_{2}$.
It should be stressed that the permutation symmetry associated with
$|\psi(j_{1}m_{1}v_{1}j_{2}m_{2}v_{2})\rangle_{\pm}$ has nothing to
with the permutation symmetry of the total wavefunction of identical particles, since
the states $|\psi(j_{1}m_{1}v_{1}j_{2}m_{2}v_{2})\rangle_{\pm}$ describe
only rotational and vibrational motion.  Note also, for use later below, that
entangled states $|\psi(j_{1}m_{1}v_{1}j_{2}m_{2}v_{2})\rangle_{\pm}$
also form a new set of basis states to describe a general entangled state, i.e.,
\begin{eqnarray}
|\psi(j_{1}m_{1}v_{1}j_{2}m_{2}v_{2})\rangle_{\alpha, \beta} &=&
\frac{\cos(\alpha)+\sin(\alpha)\exp(i\beta)}{\sqrt{2}}|\psi(j_{1}m_{1}v_{1}j_{2}m_{2}v_{2})\rangle_{+}
\nonumber
\\
&&+\frac{\cos(\alpha)-\sin(\alpha)\exp(i\beta)}{\sqrt{2}}|\psi(j_{1}m_{1}v_{1}j_{2}m_{2}v_{2})\rangle_
{-}.
\label{psiab}
\end{eqnarray}

Suppose that prior to the AB + AB collision the AB molecules (here labeled
AB and AB$^{\prime}$ for convenience) are prepared, in accord with Ref.
\cite{pauljcp}, in the superposition states:
\begin{eqnarray}
|\psi_{AB} \rangle &=& \cos(\alpha_{1})|j_{1}m_{1}v_{1}\rangle +
\sin(\alpha_{1})\exp(i\beta_{1})|j_{2}m_{2}v_{2}\rangle \nonumber \\
|\psi_{AB'} \rangle &=& \cos(\alpha_{2})|j_{1}m_{1}v_{1}\rangle  +
\sin(\alpha_{2})\exp(i\beta_{2})|j_{2}m_{2}v_{2}\rangle
\label{sups}
\end{eqnarray}
Then the total internal wavefunction (before symmetrization)  is a direct
product $|\psi_{dp}\rangle$ of these two superposition states,
rather than an entangled molecular state. Nevertheless,
$|\psi_{dp}\rangle$ can be expressed in terms of an entangled state plus
two additional components, i.e.,
\begin{eqnarray}
|\psi_{dp}\rangle&=&
y\exp(i\beta_{2})
|\psi(j_{1}m_{1}v_{1}j_{2}m_{2}v_{2})
\rangle_{\alpha\beta}
 + \cos(\alpha_{1})\cos(\alpha_{2})|j_{1}m_{1}v_{1}\rangle\otimes |j_{1}m_{1}v_{1}\rangle
 \nonumber \\ && + \sin(\alpha_{1})\sin(\alpha_{2})\exp[i(\beta_{1}+\beta_{2})]
 |j_{2}m_{2}v_{2}\rangle\otimes
 |j_{2}m_{2}v_{2}\rangle,
 \label{psidp}
 \end{eqnarray}
where the entangled state component $|\psi(j_{1}m_{1}v_{1}j_{2}m_{2}v_{2})
\rangle_{\alpha\beta}$ is given by Eq. (\ref{psiab}), with
\begin{eqnarray}
y & = &\sqrt{\cos^{2}(\alpha_{1})
\sin^{2}(\alpha_{2})+\sin^{2}(\alpha_{1})\cos^{2}(\alpha_{2})}, \nonumber \\
\alpha &= & \cos^{-1}[\cos(\alpha_{1})\sin(\alpha_{2})/y], \nonumber \\
\beta &= & \beta_{1}-\beta_{2},
\end{eqnarray}
and the two additional components $|j_{1}m_{1}v_{1}\rangle\otimes |j_{1}m_{1}v_{1}\rangle$ and
$|j_{2}m_{2}v_{2}\rangle\otimes
|j_{2}m_{2}v_{2}\rangle$ are called, in the spirit of previous
coherent control work,  ``satellite" states \cite{pauljcp}.
From this viewpoint, the contribution from the entangled state
component is the control term.

If our interest is in coherent control of AB + AB scattering, then this provides
an obvious route to control. That is, by altering the $\alpha_i$ and $\beta_i$ in the prepared
diatomic state, one can alter the initial state. This introduces
controllable quantum  interferences, as previously discussed \cite{pauljcp},
hence altering product cross sections.

Alternatively, if our interest is in the dynamics of entangled states then
collisions of AB + AB in the prescribed initial states may well provide a
route to directly observing entanglement-induced quantum effects. Specifically, there clearly are
cases, some cited in computations below, where contributions from the
entangled molecular state dominate over that of the satellite states. That
is, for some channels characterized by $j_{1}', v_{1}', j_{2}', v_{2}'$,
the scattering amplitude for the initial direct product state
$|\psi_{dp}\rangle$ is entirely due to the
$y\exp(i\beta_{2})|\psi(j_{1}m_{1}v_{1}j_{2}m_{2}v_{2})
\rangle_{\alpha\beta}$ component. For these cases,  we  have
\begin{eqnarray}
\sigma(\theta,|\psi_{dp}\rangle)\approx y^{2}
\sigma\left(\theta,|\psi(j_{1}m_{1}v_{1}j_{2}m_{2}v_{2})
_{\alpha\beta}\rangle\right),
\label{sigmaeq}
\end{eqnarray}
where $\sigma(\theta,|\psi\rangle)$ represents the differential cross
section summed over $m_{1}'$ and $m_{2}'$ and integrated over 
the azimuthal angle 
$\phi$ for
the initial state $|\psi\rangle$. Alternatively, and more difficult to
implement, 
if we
restrict the measurement to a particular total energy of the product,
contributions of the satellite states
can be zero due to the energy restriction and the scattering becomes essentially
that due to an initially entangled molecular state
$|\psi(j_{1}m_{1}v_{1}j_{2}m_{2}v_{2}) \rangle_{\alpha\beta}$. Thus,
significantly, in either case
one can experimentally observe and utilize
entanglement-assisted  coherent control effects in diatom-diatom
scattering without the difficult task of preparing entangled molecular
states.

Below we develop a general description of the quantum effects associated
with entanglement of molecular rovibrational motion in bimolecular
scattering. We focus on the contribution from the entangled state since
the remainder (i.e., the satellite terms) is ordinary scattering. In doing
so we specifically study the scattering system para--H$_2$ +  para-
H$_2$, the simplest case of diatom-diatom scattering. Although many of
the following arguments apply to both reactive and nonreactive scattering,
we confine attention to elastic and inelastic scattering only.

\section{Scattering From Entangled Rovibrational States: Formalism}
\label{formalism}

Consider AB + AB scattering. We represent the vector pointing from
the center of mass of the projectile to that of the target by ${\bf R}$, and
the relative inter-atomic distance vectors of the two para--H$_2$
molecules by ${\bf r}_{1}$ and ${\bf r}_{2}$, respectively.
We  use $j_{i}, m_{i}, v_{i}\ (i=1,2)$
to denote the
quantum numbers of angular momentum,  the projection of angular momentum
onto a space-fixed $z$ axis, and vibrational motion, respectively,
for each para--H$_2$ molecule.
The fact that both the nuclear spin and the electronic spin of para--H$_2$ molecules
are zero greatly simplifies the problem, while retaining the essence of physics.
Note that each para--H$_2$ molecule (with electronic ground state) is a boson,
so that the permutation symmetry of the total wavefunction $|\Psi({\bf r}_{1}, {\bf r}_{2},
{\bf R})\rangle$ in the scattering problem is $+1$. That is,
\begin{eqnarray}
|\Psi({\bf r}_{1}, {\bf r}_{2}, {\bf R})\rangle =
|\Psi({\bf r}_{2}, {\bf r}_{1}, -{\bf R})\rangle.
\label{boson}
\end{eqnarray}
As we shall see more clearly below,
this implies that the parity of $|\Psi({\bf r}_{1}, {\bf r}_{2},
{\bf R})\rangle $ with regard to ${\bf R}$ depends on the permutation symmetry
of
the internal degrees of freedom ${\bf r}_{1}$ and ${\bf r}_{2}$.

In traditional scattering, the initial state
is characterized by $k, \hat{{\bf z}},$ $j_{1}, m_{1}, v_{1}$, $
j_{2}, m_{2}, v_{2}$,  and the final state characterized by
$k', \hat{{\bf R}}, j_{1}'$, $m_{1}', v_{1}', j_{2}', m_{2}', v_{2}'$,
where $\hbar k$ and $\hbar k'$ are the translational 
momenta for the initial and final states, with
$k=k'$  ($k\ne k'$) in the case of elastic (inelastic) scattering.
Below we exclude the case of
$(j_{1}, m_{1}, v_{1})=(j_{2}, m_{2}, v_{2})$ as it does not
allow for entanglement in rovibrational states.
Since the colliding molecules are identical,  both the incoming and
the outgoing asymptotic states have to be appropriately symmetrized.
However, to obtain the scattering amplitude
it suffices to  symmetrize  either
the incoming or outgoing state only \cite{taylor51}.
If we choose to symmetrize the incoming state then the scattering amplitude
is given by
\begin{eqnarray}
& & f(k'\hat{{\bf R}}j_{1}'m_{1}'v_{1}'j_{2}'m_{2}'v_{2}'\leftarrow k\hat{{\bf z}}j_{1}m_{1}v_{1}
j_{2}m_{2}v_{2}) \nonumber \\
& & = \tilde{f}(k'\hat{{\bf R}}j_{1}'m_{1}'v_{1}'j_{2}'m_{2}'v_{2}'\leftarrow
k\hat{{\bf z}}j_{1}m_{1}v_{1}j_{2}m_{2}v_{2})  \nonumber  \\  &&\ \ \  +
\tilde{f}(k'\hat{{\bf R}}j_{1}'m_{1}'v_{1}'j_{2}'m_{2}'v_{2}'\leftarrow
k(-\hat{{\bf z}})j_{2}m_{2}v_{2}j_{1}m_{1}v_{1}),
\label{fexpression}
\end{eqnarray}
where $\tilde{f}$ represents the unsymmetrized scattering amplitudes under
the assumption that the two molecules
are distinguishable.  Specifically, in terms of the $T$-matrix elements, we have
\cite{green,kouri}
\begin{eqnarray}
& &\tilde{f}(k'\hat{{\bf R}}j_{1}'m_{1}'v_{1}'j_{2}'m_{2}'v_{2}'\leftarrow
k\hat{{\bf z}}j_{1}m_{1}v_{1}j_{2}m_{2}v_{2})  \nonumber \\
& & = \frac{i\sqrt{\pi}}{\sqrt{kk'}}\sum_{JMm'}\sum_{lj_{12}m_{12}}\sum_{l'j_{12}'m_{12}'}
  \sqrt{2l+1}i^{l-l'}
Y^{m'}_{l'}(\hat{{\bf R}})C^{JM}_{l'm'j_{12}'m_{12}'}C^{JM}_{l0j_{12}m_{12}} \nonumber \\
& & \ \ \ \times
C^{j_{12}'m_{12}'}_{j_{1}'m_{1}'j_{2}'m_{2}'}
C^{j_{12}m_{12}}_{j_{1}m_{1}j_{2}m_{2}}T^{JM}(j_{1}'v_{1}'j_{2}'v_{2}'j_{12}'l'|j_{1}v_{1}j_{2}v_{2}
j_{12}l),
\label{tmatrix1}
\end{eqnarray}
and
\begin{eqnarray}
& &\tilde{f}(k'\hat{{\bf R}}j_{1}'m_{1}'v_{1}'j_{2}'m_{2}'v_{2}'\leftarrow
k(-\hat{{\bf z}})j_{2}m_{2}v_{2}j_{1}m_{1}v_{1}) \nonumber \\
& & = \frac{i\sqrt{\pi}}{\sqrt{kk'}}\sum_{JMm'}\sum_{lj_{12}m_{12}}\sum_{l'j_{12}'m_{12}'}(-1)^{l}
   \sqrt{2l+1}i^{l-l'}
  Y^{m'}_{l'}(\hat{{\bf R}})C^{JM}_{l'm'j_{12}'m_{12}'}C^{JM}_{l0j_{12}m_{12}}\nonumber \\
 & &\ \ \ \times
  C^{j_{12}'m_{12}'}_{j_{1}'m_{1}'j_{2}'m_{2}'}
  C^{j_{12}m_{12}}_{j_{2}m_{2}j_{1}m_{1}}T^{JM}(j_{1}'v_{1}'j_{2}'v_{2}'j_{12
  }'l'|j_{2}v_{2}j_{1}v_{1}j_{12}l),
  \label{tmatrix2}
  \end{eqnarray}
  where $Y^{m}_{l}$ is the spherical function and
  $C^{j_{3}m_{3}}_{j_{1}m_{1}j_{2}m_{2}}$
  are the Clebsch-Gordan coefficient.

Consider now the case where the molecules are initially prepared in
an entangled state of molecular rovibrational states (as in the cross
term in Eq. (\ref{psidp})):
$\langle {\bf r}_{1}, {\bf r}_{2}|\psi(j_{1}m_{1}v_{1}j_{2}m_{2}v_{2})\rangle_{\alpha, \beta}$.
The two components of $|\psi(j_{1}m_{1}v_{1}j_{2}m_{2}v_{2})\rangle_{\alpha, \beta}$ shown in
Eq. (\ref{entangled})
are degenerate in energy. Hence  they can interfere with each other, providing
interference effects that allow for control.

To simplify matters we focus on the role of the entangled states
$|\psi(j_{1}m_{1}v_{1}j_{2}m_{2}v_{2})\rangle_{\pm}$.
Consideration of three special cases is in order:

(a) $j_{1}=j_{2}, v_{1}=v_{2}$, $m_{1}\neq m_{2}$.  Here
\begin{eqnarray}
|\psi(j_{1}m_{1}v_{1}j_{2}m_{2}v_{2})\rangle_{\pm}=\frac{1}{\sqrt{2}}
(|m_{1}\rangle|m_{2}\rangle\pm |m_{2}\rangle|m_{1}\rangle)\otimes |j_{1},v_{1}\rangle\otimes
|j_{2},v_{2}\rangle.
\label{en1}
\end{eqnarray}
In this case entanglement arises only through
molecular rotation motion  along a space fixed axis.
Case (a) is therefore totally analogous to two entangled spins [see Eq. (\ref{spinEPR})].
Indeed,
quantum effects arising from such entanglement can also be understood in terms of
polarization phenomena, in accord with general polarization theory \cite{polartheory}.

(b) $m_{1}=m_{2}, j_{1}=j_{2}$, $v_{1}\neq v_{2}$.  In this case we have
\begin{eqnarray}
|\psi(j_{1}m_{1}v_{1}j_{2}m_{2}v_{2})\rangle_{\pm}=\frac{1}{\sqrt{2}}
(|v_{1}\rangle|v_{2}\rangle\pm |v_{2}\rangle|v_{1}\rangle)\otimes
|m_{1},j_{1}\rangle\otimes
|m_{2},j_{2}\rangle.
\label{en2}
\end{eqnarray}
Here only vibrational motion is entangled;
rotational motion is completely separable.  Further, unlike
case (a),  the resulting quantum entanglement is entirely due to
internal excitation.

(c) $m_{1}=m_{2}, v_{1}=v_{2}$,
$j_{1}\neq j_{2}$. Thus,
\begin{eqnarray}
|\psi(j_{1}m_{1}v_{1}j_{2}m_{2}v_{2})\rangle_{\pm}=\frac{1}{\sqrt{2}}
(|j_{1}\rangle|j_{2}\rangle\pm |j_{2}\rangle|j_{1}\rangle)\otimes
|m_{1},v_{1}\rangle\otimes
|m_{2},v_{2}\rangle.
\label{en3}
\end{eqnarray}
This case is somewhat complicated.  Equation (\ref{en3}) implies that,
while vibrational motion is not entangled, rotational degrees of freedom
are partially entangled, insofar as the motion along a space fixed
axis is still separable.  Note that in this case quantum entanglement
is also due to internal excitation, and that it is intrinsically different
from case (a) and unrelated to
any description based on polarization theory.

Using Eqs. (\ref{fexpression}) and (\ref{definepm}),
one obtains the following scattering amplitude $f_{\pm}(\hat{{\bf R}})$ from
the initial entangled state $(k, \hat{{\bf z}}, |\psi(j_{1}m_{1}v_{1}j_{2}m_{
2}v_{2})\rangle_{\pm})$ to the final state
$k', \hat{{\bf R}}, j_{1}'$, $m_{1}', v_{1}', j_{2}', m_{2}', v_{2}'$:
\begin{eqnarray}
f_{\pm}(\hat{{\bf R}})& = &\frac{1}{\sqrt{2}}\tilde{f}(k'\hat{{\bf R}}j_{1}'m_{1}'v_{1}'j_{2}'m_{2}'v_{2}'\leftarrow
k\hat{{\bf z}}j_{1}m_{1}v_{1}j_{2}m_{2}v_{2})  \nonumber  \\ &&+
\frac{1}{\sqrt{2}}\tilde{f}(k'\hat{{\bf R}}j_{1}'m_{1}'v_{1}'j_{2}'m_{2}'v_{2}'\leftarrow
k(-\hat{{\bf z}})j_{2}m_{2}v_{2}j_{1}m_{1}v_{1}) \nonumber \\ &&\pm
\frac{1}{\sqrt{2}}\tilde{f}(k'\hat{{\bf R}}j_{1}'m_{1}'v_{1}'j_{2}'m_{2}'v_{2}'\leftarrow
k\hat{{\bf z}}j_{2}m_{2}v_{2}j_{1}m_{1}v_{1}) \nonumber \\ &&\pm
\frac{1}{\sqrt{2}}\tilde{f}(k'\hat{{\bf R}}j_{1}'m_{1}'v_{1}'j_{2}'m_{2}'v_{2}'\leftarrow
k(-\hat{{\bf z}})j_{1}m_{1}v_{1}j_{2}m_{2}v_{2}).
\label{fpm}
\end{eqnarray}
Integrating  $|f_{\pm}(\hat{{\bf R}})|^2$ over $\hat{{\bf R}}$ gives
$|f_{\pm}|^2$, which is proportional to the cross section for the
$j_{1}'m_{1}'v_{1}'j_{2}'m_{2}'v_{2}'\leftarrow
j_{1}m_{1}v_{1}j_{2}m_{2}v_{2}$ transition, with the initial state given
by the plus or minus combination. The square of the amplitude for the same
transition integrated over $\hat{{\bf R}}$, but from  the collision where the initial state is
AB($j_{1}m_{1}v_{1}$) + AB($j_{2}m_{2}v_{2}$) [properly symmetrized] is
given by $|f|^2$. Then, using  Eqs. (\ref{fexpression}) 
and (\ref{fpm}) we have that
\begin{equation}
|f|^2 = \frac{|f_+|^2 + |f_-|^2}{2}.
\end{equation}
By comparing $|f_+|^2$, $|f_-|^2$ and $|f|^2$ we can ascertain the effect
of entangling rovibrational states, and ascertain the degree of control.

Given Eqs. (\ref{tmatrix1}), (\ref{tmatrix2}) and (\ref{fpm}),
the $f_{\pm}(\hat{{\bf R}}) $ can be written as:
\begin{eqnarray}
f_{\pm}(\hat{{\bf R}})&=&\frac{1}{\sqrt{2}}
\frac{i\sqrt{\pi}}{\sqrt{kk'}}\sum_{JMm'}\sum_{lj_{12}m_{12}}\sum_{l'j_{12}'m_{12}'}
 \sqrt{2l+1}i^{l-l'} Y^{m'}_{l'}(\hat{{\bf R}})
  C^{JM}_{l'm'j_{12}'m_{12}'}C^{JM}_{l0j_{12}m_{12}}
  C^{j_{12}'m_{12}'}_{j_{1}'m_{1}'j_{2}'m_{2}'} \nonumber \\
  && \times\{ [1 \pm (-1)^{l}] [
 C^{j_{12}m_{12}}_{j_{1}m_{1}j_{2}m_{2}}T^{JM}(j_{1}'v_{1}'j_{2}'v_{2}'j_{12}'l'|j_{1}v_{1}j_{2}v_{2}
 j_{12}l)  \nonumber \\
& &\pm
C^{j_{12}m_{12}}_{j_{2}m_{2}j_{1}m_{1}}
T^{JM}(j_{1}'v_{1}'j_{2}'v_{2}'j_{12}'l'|j_{2}v_{2}j_{1}v_{1}j_{12}l)]\}.
\label{ftmatrix}
\end{eqnarray}

Equation (\ref{ftmatrix}) suggests that there are  two important quantum
effects in AB + AB scattering associated with using entangled molecular
states as initial states: (1) Due to the factor $[1 \pm (-1)^{l}]$, the
permutation symmetry induced by molecular entanglement imposes a parity
restriction on the incoming partial waves. That is, for the case of
$|\psi(j_{1}m_{1}v_{1}j_{2}m_{2}v_{2})\rangle_{+}$ ($|\psi(
j_{1}m_{1}v_{1}j_{2}m_{2}v_{2})\rangle_{-}$),  contributions from odd
(even) partial waves are completely suppressed, whereas contributions from
even (odd) partial waves are enhanced. For example, for the case of
$|\psi( j_{1}m_{1}v_{1}j_{2}m_{2}v_{2})\rangle_{-}$, although the two
para--H$_2$ molecules are spinless bosons, they avoid the $l=0$ state
due to quantum entanglement of rovibrational motion. (2) As indicated by
the factor
$[C^{j_{12}m_{12}}_{j_{1}m_{1}j_{2}m_{2}}T^{JM}(j_{1}'v_{1}'j_{2}'v_{2}'j_{12}'l'|j_{1}v_{1}j_{2}v_{2}
 j_{12}l)\pm C^{j_{12}m_{12}}_{j_{2}m_{2}j_{1}m_{1}}
 T^{JM}(j_{1}'v_{1}'j_{2}'v_{2}'j_{12}'l'|j_{2}v_{2}j_{1}v_{1}j_{12}l)]$ in Eq. (\ref{ftmatrix}),
 there is  quantum interference  between
transitions $(j_{1}v_{1}j_{2}v_{2}j_{12}l)$ $\rightarrow
(j_{1}'v_{1}'j_{2}'v_{2}'j_{12}'l') $ and $(j_{2}v_{2}j_{1}v_{1}j_{12}l)$
$\rightarrow (j_{1}'v_{1}'j_{2}'v_{2}'j_{12}'l') $.  Whether this
interference is constructive or destructive depends on the form of the
entangled rovibrational state. This effect is significant when the
magnitude of
$T^{JM}(j_{1}'v_{1}'j_{2}'v_{2}'j_{12}'l'|j_{1}v_{1}j_{2}v_{2}j_{12}l)$ is
comparable to that of
$T^{JM}(j_{1}'v_{1}'j_{2}'v_{2}'j_{12}'l'|j_{2}v_{2}j_{1}v_{1}j_{12}l)$.
Further, based on Eq. (\ref{ftmatrix}), the scattering amplitude
$f_{\alpha, \beta}(\hat{{\bf R}})$ for an arbitrary incoming entangled state $(k,
\hat{{\bf z}}, |\psi(j_{1}m_{1}v_{1}j_{2}m_{2}v_{2})\rangle_{\alpha,
\beta})$ can then  be expressed as
\begin{eqnarray}
f_{\alpha,\beta}(\hat{{\bf R}})=\frac{\cos(\alpha)+\sin(\alpha)\exp(i\beta)}{\sqrt{2}}
f_{+}(\hat{{\bf
R}})
+ \frac{\cos(\alpha)-\sin(\alpha)\exp(i\beta)}{\sqrt{2}}f_{-}(\hat{{\bf R}}).
\label{fab}
\end{eqnarray}

As mentioned above,  an alternative but equivalent way to
symmetrize the scattering amplitude is to symmetrize the outgoing asymptotic state.
That is,
    \begin{eqnarray}
      & & f(k'\hat{{\bf R}}j_{1}'m_{1}'v_{1}'j_{2}'m_{2}'v_{2}'\leftarrow k\hat{{\bf
    z}}j_{1}m_{1}v_{1}
      j_{2}m_{2}v_{2}) \nonumber \\
        && = \tilde{f}(k'\hat{{\bf R}}j_{1}'m_{1}'v_{1}'j_{2}'m_{2}'v_{2}'\leftarrow
         k\hat{{\bf z}}j_{1}m_{1}v_{1}j_{2}m_{2}v_{2})  \nonumber  \\ &&\ \ \  +
           \tilde{f}(k'(-\hat{{\bf
           R}})j_{2}'m_{2}'v_{2}'j_{1}'m_{1}'v_{1}'\leftarrow
     k\hat{{\bf z}}j_{1}m_{1}v_{1}j_{2}m_{2}v_{2}).
       \end{eqnarray}
Based on this procedure,  we have
\begin{eqnarray}
 f_{\pm}(\hat{{\bf R}})& = &\frac{1}{\sqrt{2}}\tilde{f}(k'\hat{{\bf
 R}}j_{1}'m_{1}'v_{1}'j_{2}'m_
 {2}'v_{2}'\leftarrow
 k\hat{{\bf z}}j_{1}m_{1}v_{1}j_{2}m_{2}v_{2})  \nonumber  \\ &&+
 \frac{1}{\sqrt{2}}\tilde{f}(k'(-\hat{{\bf
 R}})j_{2}'m_{2}'v_{2}'j_{1}'m_{1}'v_{1}'\leftarrow
 k\hat{{\bf z}}j_{1}m_{1}v_{1}j_{2}m_{2}v_{2}) \nonumber \\ &&\pm
 \frac{1}{\sqrt{2}}\tilde{f}(k'\hat{{\bf
 R}}j_{1}'m_{1}'v_{1}'j_{2}'m_{2}'v_{2}'\leftarrow
 k\hat{{\bf z}}j_{2}m_{2}v_{2}j_{1}m_{1}v_{1}) \nonumber \\ &&\pm
 \frac{1}{\sqrt{2}}\tilde{f}(k'(-\hat{{\bf
 R}})j_{2}'m_{2}'v_{2}'j_{1}'m_{1}'v_{1}'\leftarrow
 k\hat{{\bf z}}j_{2}m_{2}v_{2}j_{1}m_{1}v_{1}).
 \label{fpm2}
 \end{eqnarray}
In terms of the partial wave expansion,
the scattering amplitude $ f_{\pm}(\hat{{\bf R}})$ in Eq. (\ref{fpm2}) can be
further expressed
as
\begin{eqnarray}
f_{\pm}(\hat{{\bf R}})&=&\frac{1}{\sqrt{2}}
\frac{i\sqrt{\pi}}{\sqrt{kk'}}\sum_{JMm'}\sum_{lj_{12}m_{12}}\sum_{l'j_{12}'m_{12}'}
 \sqrt{2l+1}i^{l-l'}
 C^{JM}_{l'm'j_{12}'m_{12}'}C^{JM}_{l0j_{12}m_{12}}
  \nonumber \\
   & & \times  \{C^{j_{12}m_{12}}_{j_{1}m_{1}j_{2}m_{2}} [C^{j_{12}'m_{12}'}_{j_{1}'m_{1}'j_{2}'m_{2}'}
   Y^{m'}_{l'}(\hat{{\bf R}})T^{JM}(j_{1}'v_{1}'j_{2}'v_{2}'j_{12}'l'|j_{1}v_{1}j_{2}v_{2}
      j_{12}l) \nonumber \\
    &&+
   C^{j_{12}'m_{12}'}_{j_{2}'m_{2}'j_{1}'m_{1}'}Y^{m'}_{l'}(-\hat{{\bf R}})
   T^{JM}(j_{2}'v_{2}'j_{1}'v_{1}'j_{12}'l'|j_{1}v_{1}j_{2}v_{2}
    j_{12}l)]  \nonumber \\
    &&\pm
    C^{j_{12}m_{12}}_{j_{2}m_{2}j_{1}m_{1}}[C^{j_{12}'m_{12}'}_{j_{1}'m_{1}'j_{2}'m_{2}'}Y^{
    m'}_{l'}(\hat{{\bf R}})T^{JM}(j_{1}'v_{1}'j_{2}'v_{2}'j_{12}'l'|j_{2}v_{2}j_{1}v_{1}j_{12}l) \nonumber \\
   &&\pm
   C^{j_{12}'m_{12}'}_{j_{2}'m_{2}'j_{1}'m_{1}'}Y^{m'}_{l'}(-\hat{{\bf R}})
    T^{JM}(j_{2}'v_{2}'j_{1}'v_{1}'j_{12}'l'|j_{2}v_{2}j_{1}v_{1}j_{12}l)]\}.
    \label{ftmatrix2}
    \end{eqnarray}
Interestingly,  the equivalence of Eq. (\ref{ftmatrix2}) with Eq.
(\ref{ftmatrix}) is far from obvious \cite{parity-note}. As a result, we
can no longer readily identify the parity restriction condition imposed on
the incoming partial waves due to molecular entanglement. In this sense,
in the theoretical considerations, we prefer Eq. (\ref{ftmatrix}) to Eq.
(\ref{ftmatrix2}). However,  Eq. (\ref{ftmatrix2}) is very useful in order
to confirm the consistency of our results. For all computational examples
presented below we have verified that Eqs. (\ref{ftmatrix}) and 
(\ref{ftmatrix2}) give the same results.

\section{Calculations and Results}
\label{s5-3}

In this section we focus on computations for case (c) described above. Specifically,
we consider rotational energy transfer in rotor-rotor scattering, a problem
that has been the subject of considerable interest \cite{green,kouri}. We
do not treat case (a) since it is an example of well-known polarization theory.
Case (b), on the other hand, requires computations beyond current 
capabilities. However, to provide some insight into this case we do
present results for one case (b) {\em model}. That is, we treat a lower
dimensional case of vibrational excitation with frozen rotational motion.

MOLSCAT \cite{molscat} is a useful tool for this study. In particular, it
provides both cross sections and $S$-matrix elements necessary for the
study of phase control. MOLSCAT also introduces a control parameter to
account for whether or not the scattering particles are identical. Thus,
we can first assume that the two scattering molecules are distinguishable
and compute all the matrix elements
$T^{JM}(j_{1}'v_{1}'j_{2}'v_{2}'j_{12}'l'|j_{1}v_{1}j_{2}v_{2}j_{12}l)$
and
$T^{JM}(j_{1}'v_{1}'j_{2}'v_{2}'j_{12}'l'|j_{2}v_{2}j_{1}v_{1}j_{12}l)$,
and then use Eq. (\ref{ftmatrix}) to calculate the cross sections. The
Zarur-Rabitz interaction potential \cite{Rabitz2} for para--H$_2$ +
para--H$_2$ is used throughout; it certainly suffices for the
demonstration of novel quantum effects in bimolecular scattering. However,
newer global potentials are available for $H_2$ + $H_2$
scattering \cite{4hpotential} that can be used for detailed comparisons
with future experimental studies.

\subsection{Scattering at $E_{k}=400, 40, 4 $  cm$^{-1}$}

To demonstrate the role of entanglement we consider scattering for
different incoming entangled molecular states at collision energies
$E_{k}=$ 400, 40, and 4 cm$^{-1}$.  In particular, $E_{k}=400$ cm$^{-1}$
($\approx 570^o$ K) is still below the vibration excitation threshold, and
$E_{k}=4$ cm$^{-1}$ ($\approx 5.7^o$ K) is very close to the lowest
collision temperature  currently achievable in  molecular crossed-beam
experiments.

We consider the following entangled molecular states in our computations:
\begin{eqnarray}
|\psi_{j_{1}j_{2}}^{\pm}\rangle=
\frac{1}{\sqrt{2}}(|j_{1}\rangle |j_{2}\rangle \pm |j_{2}\rangle |j_{1}\rangle)
\otimes |m_{1}=0,v_{1}=0\rangle \otimes |m_{2}=0,v_{2}=0\rangle.
\label{j1j2}
\end{eqnarray}
Subsequent computational investigation \cite{vlado} of other $\alpha,
\beta$ values in Eq. (\ref{psiab}) confirmed that these two states give the
control extremes.

Since we consider rotational excitation only, all vibrational quantum
numbers are set to be zero, i.e., $v_{1}=v_{2}=v_{1}'=v_{2}'=0.$ In a
typical experiment one might well prepare the AB diatomic superposition
state by laser excitation of $|j_1,m_1,v_1 \rangle$ to $|j_2,m_2,v_2
\rangle$. Assuming linearly polarized light, selection rules ensure that
$m_1 = m_2$. However, as long as $m_{1}=m_{2}$ (so that we can not
distinguish between the two scattering molecules by measuring the
projection of their angular momentum), the results for all $m$ are
essentially the same. Hence, here we consider $m_1=m_2=0$.

\begin{figure}[ht]
\begin{center}
\epsfig{file=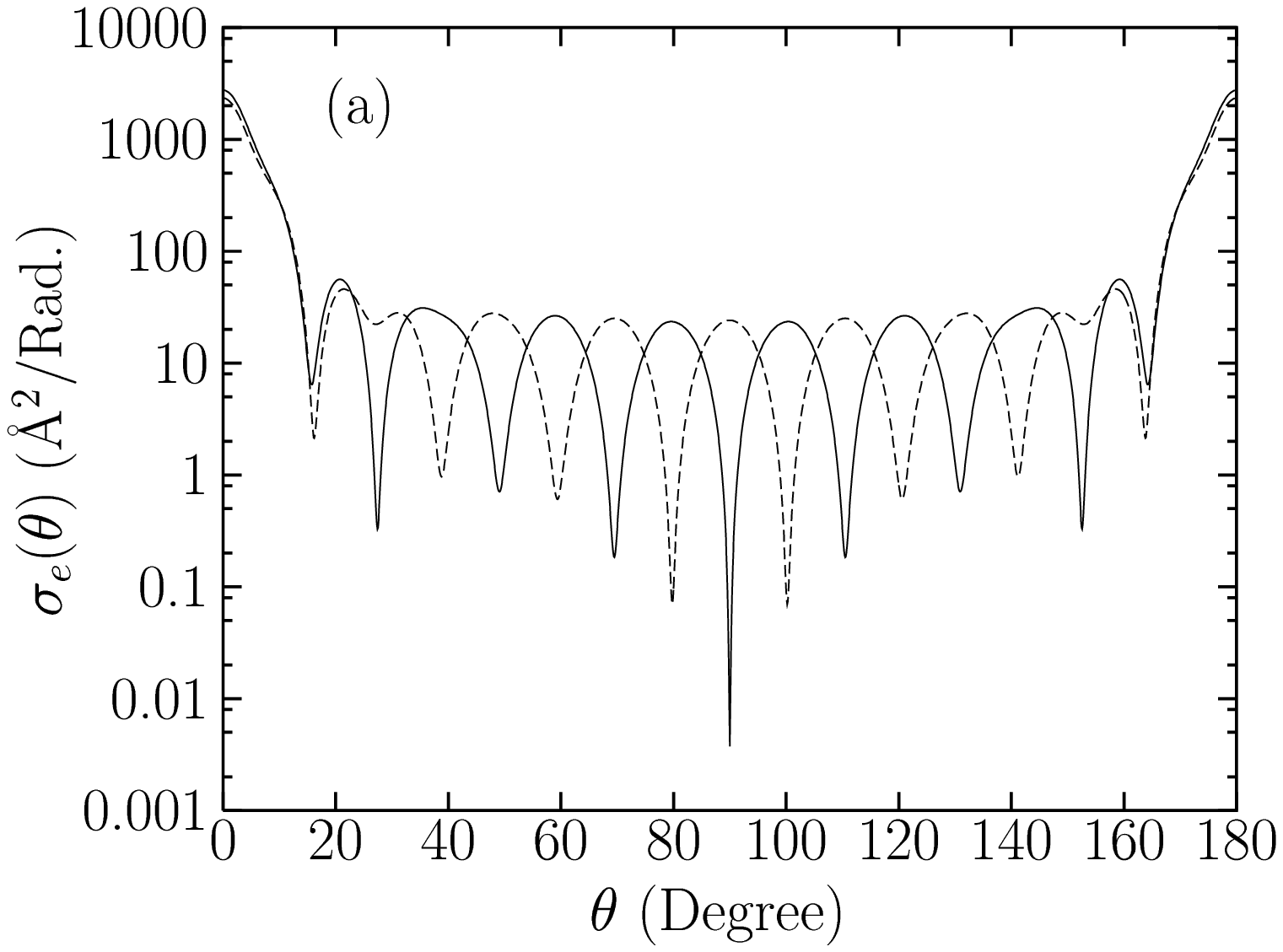,width=6.5cm}

\vspace{-1.4cm}
\epsfig{file=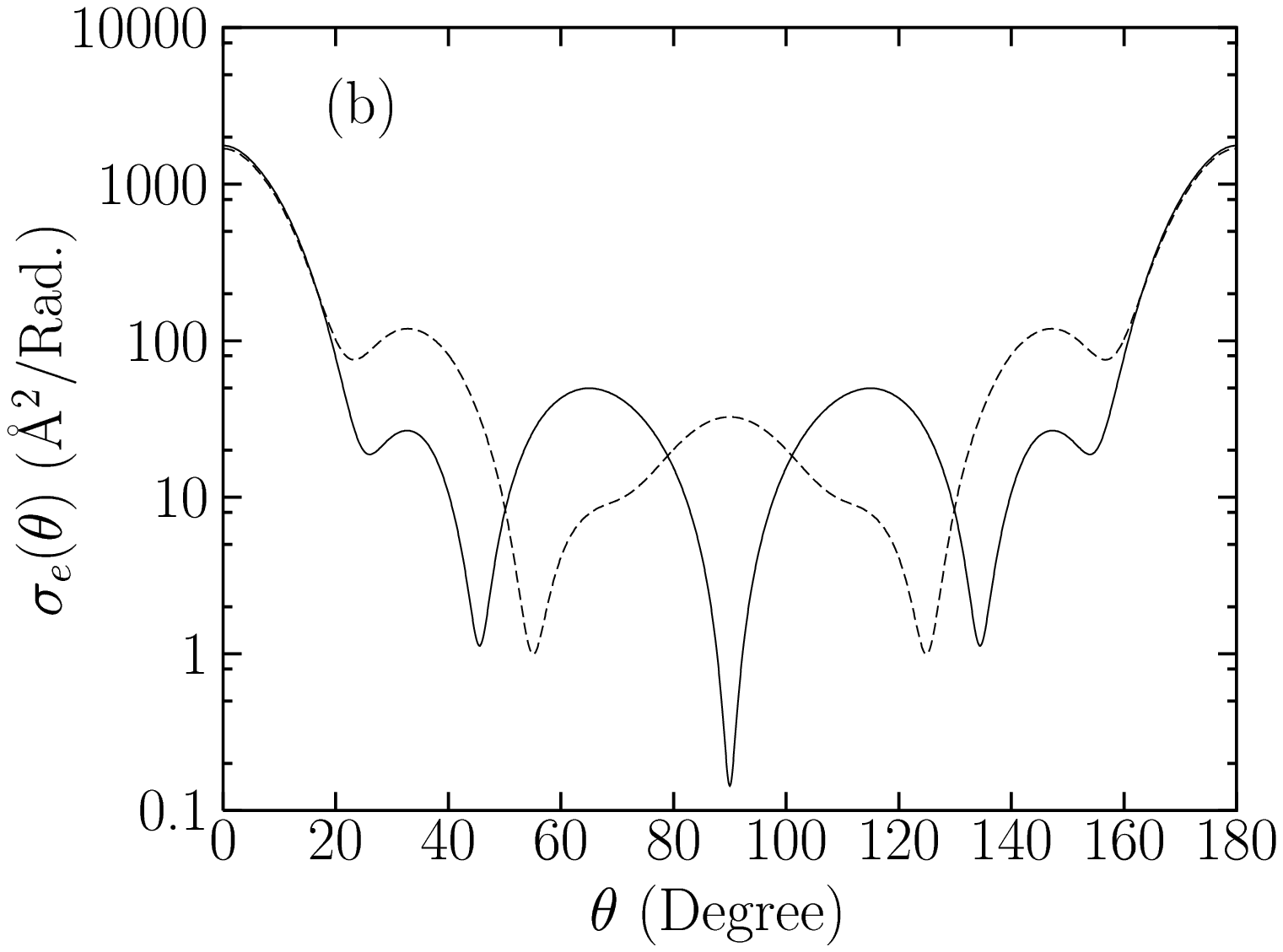,width=6.5cm}

\vspace{-1.4cm}
\epsfig{file=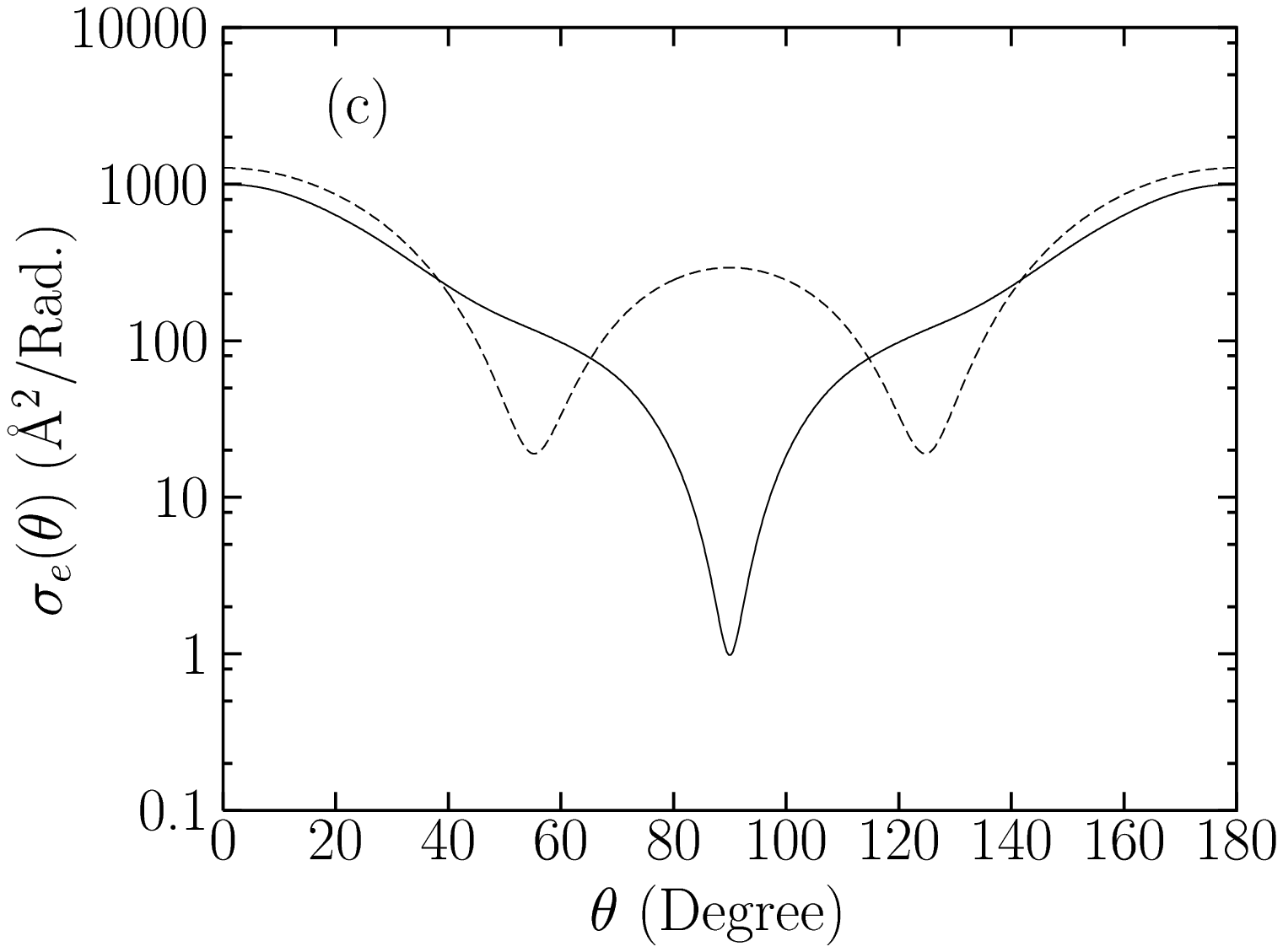,width=6.5cm}
\end{center}
\vspace{-1cm}
\caption{The $\theta$ dependence of the elastic differential cross section summed
over the final state quantum numbers $m_{1}'$ and $ m_{2}'$ and integrated
over $\phi$. The scattering system is para--H$_2$ $+$ para--H$_2$, and
the collision energy is (a) 400 cm$^{-1}$ ($\approx 570^o$ K), (b) $40$
cm$^{-1}$ ($\approx 57^o$ K), and (c) $4$ cm$^{-1}$ ($\approx 5.7^o$ K).
Dashed and solid lines are for the incoming free entangled states
$|\psi_{j_{1}j_{2}}^{+}\rangle$ and $|\psi_{j_{1}j_{2}}^{-}\rangle$ [see
Eq. (\ref{j1j2})], respectively. Here $j_{1}=2$, $ j_{2}=0$.
}
\label{Fig5-1}
\end{figure}

Figure \ref{Fig5-1} shows the $\theta$ dependence of the elastic
differential cross section $\sigma_{e}(\theta)$ summed over the
final-state quantum numbers $m_{1}'$ and $m_{2}'$ and integrated over the
azimuthal angle $\phi$ ($\hat{{\bf R}}\equiv(\theta, \phi$)), for the two
incoming entangled states $|\psi_{j_{1}j_{2}}^{\pm}\rangle$ with $j_{1}=2,
j_{2}=0$.   In particular, $E_{k}$ equals $400$ cm$^{-1}$ in Fig.
\ref{Fig5-1}a, $40$ cm$^{-1}$ in Fig. \ref{Fig5-1}b, and $4$ cm$^{-1}$ in
Fig. \ref{Fig5-1}c. For all three cases  there is a clear difference in
$\sigma_{e}(\theta)$ between the $|\psi_{j_{1}j_{2}}^{+}\rangle$ and
$|\psi_{j_{1}j_{2}}^{-}\rangle$ cases . Indeed, for particular scattering
angles, e.g., for $\theta$ close to $\pi/2$, molecular entanglement
induces huge differences. Also evident is that the number of minima of
$\sigma_{e}(\theta)$ is always even (odd) for
$|\psi_{j_{1}j_{2}}^{+}\rangle$ ($|\psi_{j_{1}j_{2}}^{-}\rangle$), a
manifestation of the parity restriction of the incoming partial waves
resulting from the permutation symmetry of the entangled molecular states.
Further, comparing Fig. \ref{Fig5-1}c with Fig. \ref{Fig5-1}a and Fig.
\ref{Fig5-1}b, it is clear that lower collision energy induces slower
oscillations in $\sigma_{e}(\theta)$.  This is expected due to the
behavior of  $Y_{l'}^{m'}(\hat{{\bf R}})$ in Eq. (\ref{ftmatrix}).

Consider then the total cross section $\sigma_{e}$ for elastic scattering,
denoted $\sigma_e^{\pm}$ for the $|\psi_{j_{1}j_{2}}^{\pm}\rangle$ cases.
Integrating $\sigma_{e}(\theta)$ over $\theta$ for the lowest energy case
(Fig. 1c) gives $\sigma_{e}^+=511$ \AA$^{2}$  and $\sigma_e^-=346$
\AA$^{2}$. This is to be compared to the total elastic cross section for
scattering of AB($j=0,m=0$) + AB ($j=2,m=0$), proportional to $|f|^2$ summed over
$m_{1}'$ and $m_{2}'$, denoted
$\sigma_e$ and equal to 428 \AA$^2$. If we define the percentage of
control in this case as $d_c = |100(\sigma^+_e - \sigma^-_e)/\sigma_e|$
then $d_c=39\%$.  For the other two higher energy cases (see Fig.
\ref{Fig5-1}a and Fig. \ref{Fig5-1}b) the cross sections are found to be less
sensitive to the quantum entanglement in the incoming state, with $d_c
=5\%$ and 13\%, respectively. Note also that the probability of the direct
transition $j_{1}=2, j_{2}=0\rightarrow j_{1}'=2, j_{2}'=0 $ is much
larger than that of the exchange transition $j_{1}=0, j_{2}=2\rightarrow
j_{1}'=2, j_{2}'=0 $. Thus, the effect from the quantum interference
between the scattering amplitudes $T^{JM}(20002l'|20002l)$ and
$T^{JM}(20002l'|00202l)$ [see Eq. (\ref{ftmatrix})] is negligible. As a
result, quantum features observed in Fig. \ref{Fig5-1} are entirely due to
the first quantum effect identified in the previous section, i.e., the
partial-wave parity selection effect.

\begin{figure}[ht]
\begin{center}
\epsfig{file=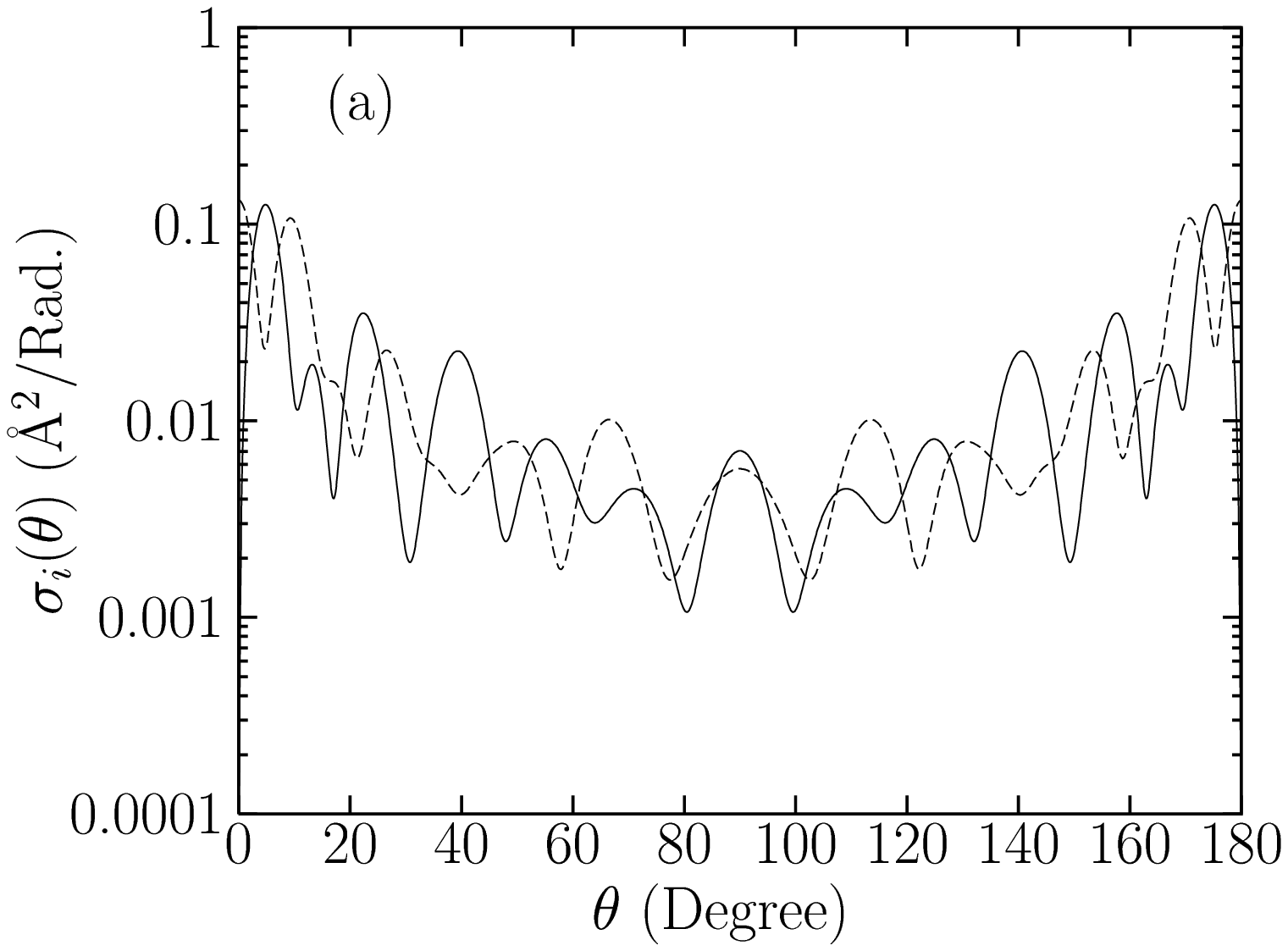,width=6.5cm}

\vspace{-1.4cm}\epsfig{file=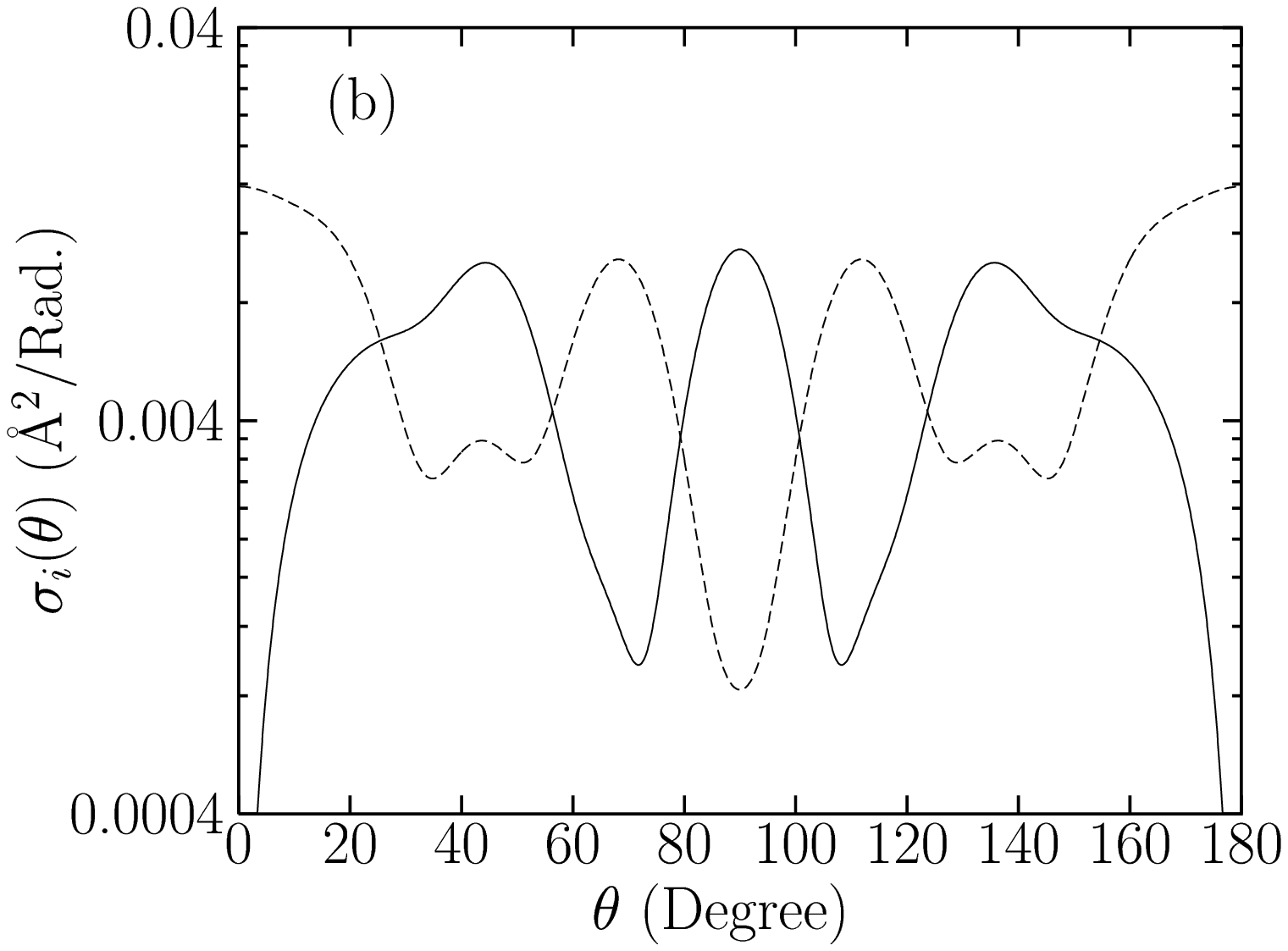,width=6.5cm}

\vspace{-1.4cm}\epsfig{file=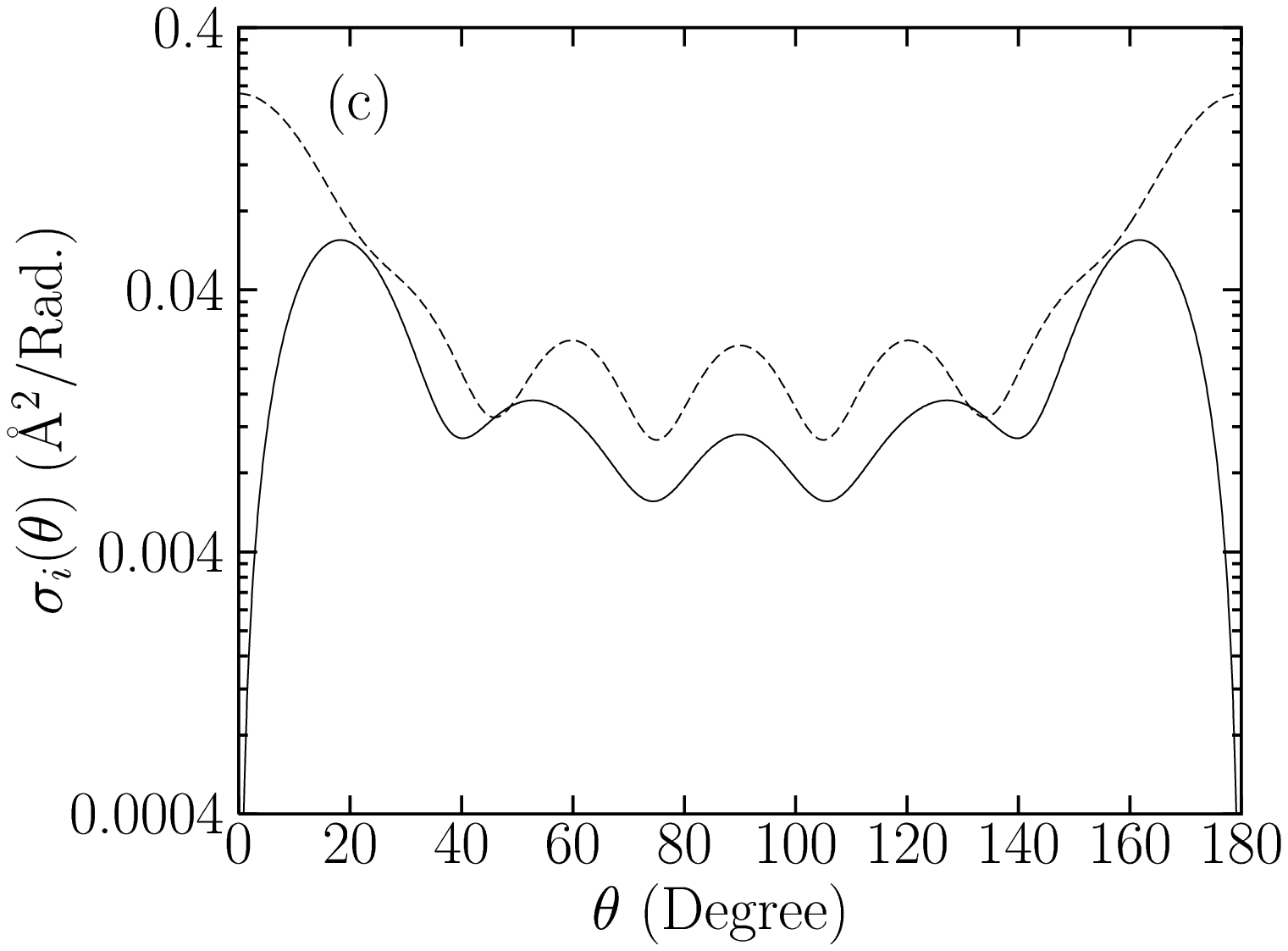,width=6.5cm}
\end{center}
\vspace{-1cm}
\caption{The $\theta$ dependence of the inelastic differential cross section summed
over the final state quantum numbers $m_{1}'$ and $ m_{2}'$ and integrated
over $\phi$. The scattering system is para--H$_{2} $+ para--H$_2$,
and the collision energy is (a) $400$ cm$^{-1}$, (b) $40$ cm$^{-1}$, and
(c) $4$ cm$^{-1}$. Dashed and solid lines are for the incoming free
entangled states $|\psi_{j_{1}j_{2}}^{+}\rangle$ and
$|\psi_{j_{1}j_{2}}^{-}\rangle$ [see Eq. (\ref{j1j2})], respectively. Here
$j_{1}=4$, $j_{2}=0$, $j_{1}'=j_{2}'=2$.
}
\label{Fig5-2}
\end{figure}

In Fig. \ref{Fig5-2}, we show inelastic differential cross sections
$\sigma_{i}(\theta)$ summed over $m_{1}'$ and $m_{2}'$ and integrated over
the angle $\phi$, for the incoming entangled states
$|\psi_{j_{1}j_{2}}^{\pm}\rangle$, with initial-state  quantum numbers
$j_{1}=4$, $j_{2}=0$, and final-state quantum numbers $j_{1}'=j_{2}'=2$.
The collision energy $E_{k}$ equals $400$ cm$^{-1}$ in Fig. \ref{Fig5-2}a,
$40$ cm$^{-1}$ in Fig. \ref{Fig5-2}b, and $4$ cm$^{-1}$ in Fig.
\ref{Fig5-2}c. For these initial-state and final-state channels, the
$T$-matrix elements contributing to the cross sections are
$T^{JM}(2020j_{12}'l'|40004l)$ and $T^{JM}(2020j_{12}'l'|00404l)$ [see Eq.
(\ref{ftmatrix})]. A careful examination of these $T$-matrix elements
shows that
\begin{eqnarray}
T^{JM}(2020j_{12}'l'|40004l)=(-1)^{j_{12}'}T^{JM}(2020j_{12}'l'|00404l).
\label{j12eq}
\end{eqnarray}
Hence, the second quantum effect described by Eq. (\ref{ftmatrix}), i.e.,
quantum interference between $T^{JM}(2020j_{12}'l'|40004l)$ and
$T^{JM}(2020j_{12}'l'|00404l)$, also becomes important.  Indeed, our
choice of the final-state channel as $j_1'=j_2'$ is motivated by the fact
that for $j_{1}'\neq j_{2}'$, the second quantum effect is insignificant.
As seen in Fig. \ref{Fig5-2}, different incoming entangled states cause
different oscillatory $\sigma_{i}(\theta)$ patterns, leading to large
differences in $\sigma_{i}(\theta)$ for fixed scattering angle (e.g., for
$\theta$ close to $0$ or $\pi$ in Fig. \ref{Fig5-2}b and Fig.
\ref{Fig5-2}c, and for $\theta$ around $\pi/2$ in Fig. \ref{Fig5-2}b). In
comparing Figs. \ref{Fig5-2}a, \ref{Fig5-2}b and \ref{Fig5-2}c, one sees
that in the highest collision energy case, quantum effects are least
significant. This trend is similar to the elastic scattering case, and can
also be understood via Eq. (\ref{ftmatrix}). In particular, using Eq.
(\ref{j12eq}) and the fact that the total parity is always conserved, Eq.
(\ref{ftmatrix}) can be further reduced to
\begin{eqnarray}
f_{\pm}(\hat{{\bf R}})&=&\frac{1}{\sqrt{2}}
\frac{i\sqrt{\pi}}{\sqrt{kk'}}\sum_{Jnlj_{12}'m_{12}'}
 \sqrt{2l+1}(-1)^{n} Y^{-m_{12}'}_{l+2n}(\hat{{\bf R}})
  C^{J0}_{(l+2n)(-m_{12}')j_{12}'m_{12}'}C^{J0}_{l040}
  C^{j_{12}'m_{12}'}_{2m_{1}'2m_{2}'} \nonumber \\
   && \times
C^{40}_{4000}[1\pm (-1)^{l}]T^{J0}(2020j_{12}'(l+2n)|40004l)[1+(-1)^{l+j_{12}'}],
\label{newftmatrix}
\end{eqnarray}
where $2n=l'-l$.  Equations (\ref{j12eq}) and (\ref{newftmatrix}) show
that quantum interference between $T^{J0}(2020j_{12}'l'|40004l)$ and
$T^{J0}(2020j_{12}'l'|00404l)$ can also be interpreted as a selection rule
imposed on the quantum number $j_{12}'$: for the incoming entangled state
$|\psi_{40}^{+}\rangle$ only even incoming partial waves contribute and
$j_{12}'$ is restricted  to be even, whereas for the incoming entangled
state $|\psi_{40}^{-}\rangle$, only odd incoming partial waves contribute
and $j_{12}'$ is restricted to be odd. This is consistent with the fact
that the outgoing state is composed of two indistinguishable molecules,
with the same rotational quantum number ($j_{1}'=j_{2}'=2$).

The case in Fig. \ref{Fig5-2}c is worthy of further discussion. In this
case one sees that $\sigma_{i}(\theta)$ for
$|\psi_{j_{1}j_{2}}^{-}\rangle$  (solid curve) is systematically smaller
than that for $|\psi_{j_{1}j_{2}}^{+}\rangle$ (dashed curve), for almost
the entire range of the scattering angle.   This leads to a significant
difference in the total inelastic cross section $\sigma_{i}^{\pm}$ for the
$|\psi_{j_{1}j_{2}}^{\pm}\rangle$ cases:
$\sigma_{i}^+= 0.057$ \AA$^{2}$ and $\sigma_i^-=0.032$ \AA$^{2}$, giving a
control range of $d_c=|100(\sigma_{i}^{+}-\sigma_{i}^{-})/\sigma_{i})|=$57\%,
where $\sigma_{i}$ is the total inelastic cross section
for scattering from AB($j=4,m=0$) + AB ($j=0,m=0$) to the channel
$j_{1}'=j_{2}'=2$.
However, $d_c$ is less than 2\% at the higher
energies in Fig. \ref{Fig5-2}. Thus, as in elastic scattering, total
inelastic cross sections
can also be sensitive to the quantum phase embedded in initial entangled
molecular states at sufficiently low energies.

\subsection{Ultracold Molecular Collisions: Extreme Quantum Effects}

Given the results for $E_{k}>5^o$ K, we consider ultracold collisions. This
new area of focus, ultracold molecular scattering, has already displayed
interesting quantum effects \cite{dalgarno}. Further,  understanding this
area is also crucial in realizing molecular Bose-Einstein condensation
\cite{superchem,bohn}.

From the control point of view, the advantage of ultracold collisions is
that scattering cross sections are often dominated by very few partial
waves. This feature of ultracold collisions may well allow for new
opportunities in manipulating molecular collisions \cite{zare}. For
example, the fact that entangled incoming molecular states impose parity
restrictions on the partial waves [see Eq. (\ref{ftmatrix})] implies that
quantum entanglement should be even more important at ultracold
temperatures. Consider, for example, a zero collision energy limit case
where only s-wave scattering ($l=0$ partial wave) contributes to the
cross sections \cite{note2}. In this case, if s-wave scattering is
selected by molecular entanglement, the result would be enhanced nonzero
cross sections. If s-wave scattering is forbidden due to molecular
entanglement, then molecular scattering is completely suppressed. That is,
in the elastic scattering case, all the scattering would be in the forward
direction, and in the completely suppressed inelastic scattering case,
there would be no inelastic scattering in any direction.

To confirm such theoretical considerations and to determine how cold the
collisions need be in order to experimentally observe such extreme quantum
effects, we extend our calculations to the ultracold regime, with $E_{k}$
ranging from $0.4 $ cm$^{-1}$ to $0.0004$ cm$^{-1}$. These computations
are generally easier than at higher energies because the maximum total
angular momentum contributing to cross sections is very small \cite{note3}.

\begin{figure}[ht]
\begin{center}
\epsfig{file=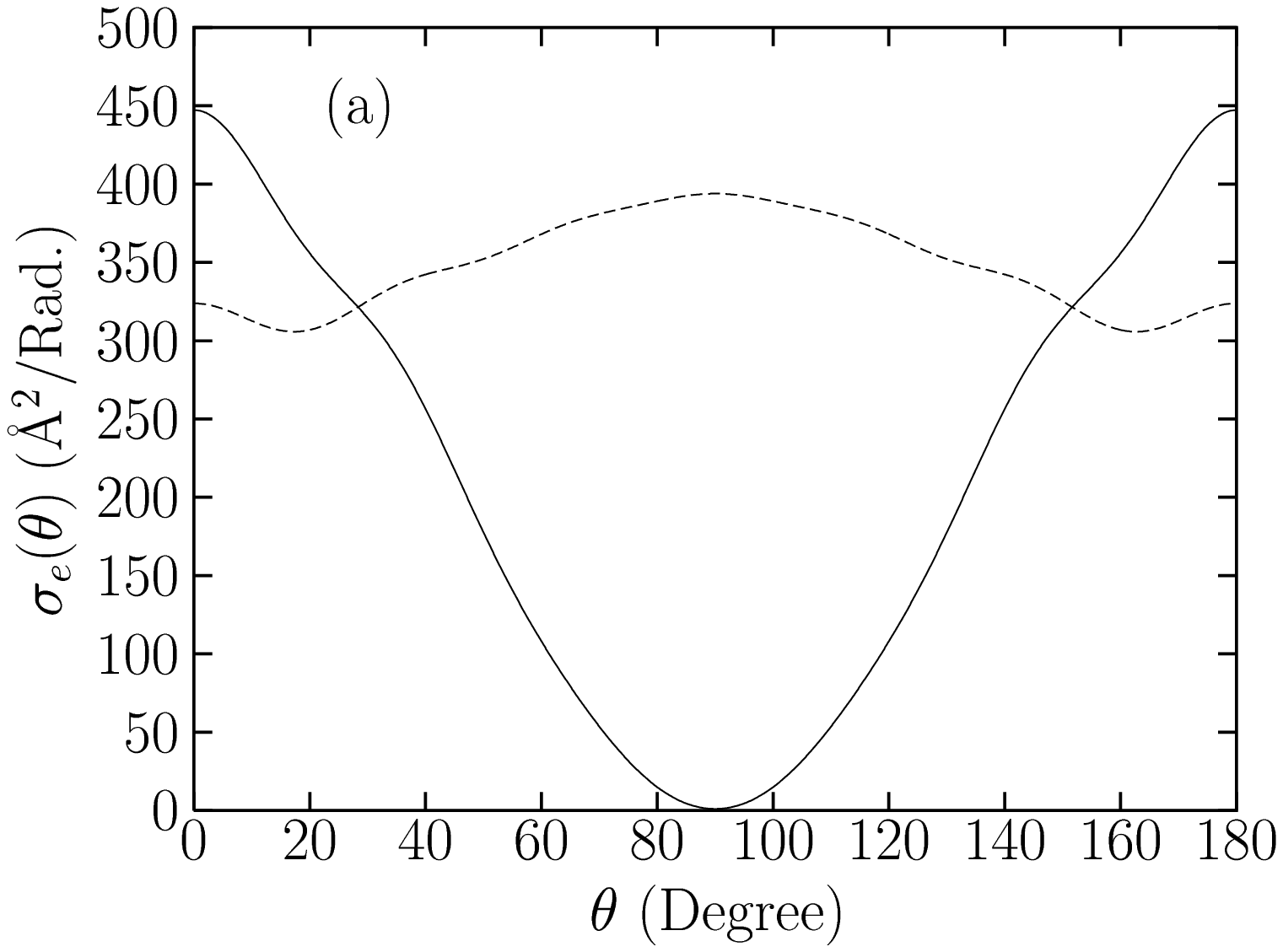,width=6.5cm}

\vspace{-1.4cm}\epsfig{file=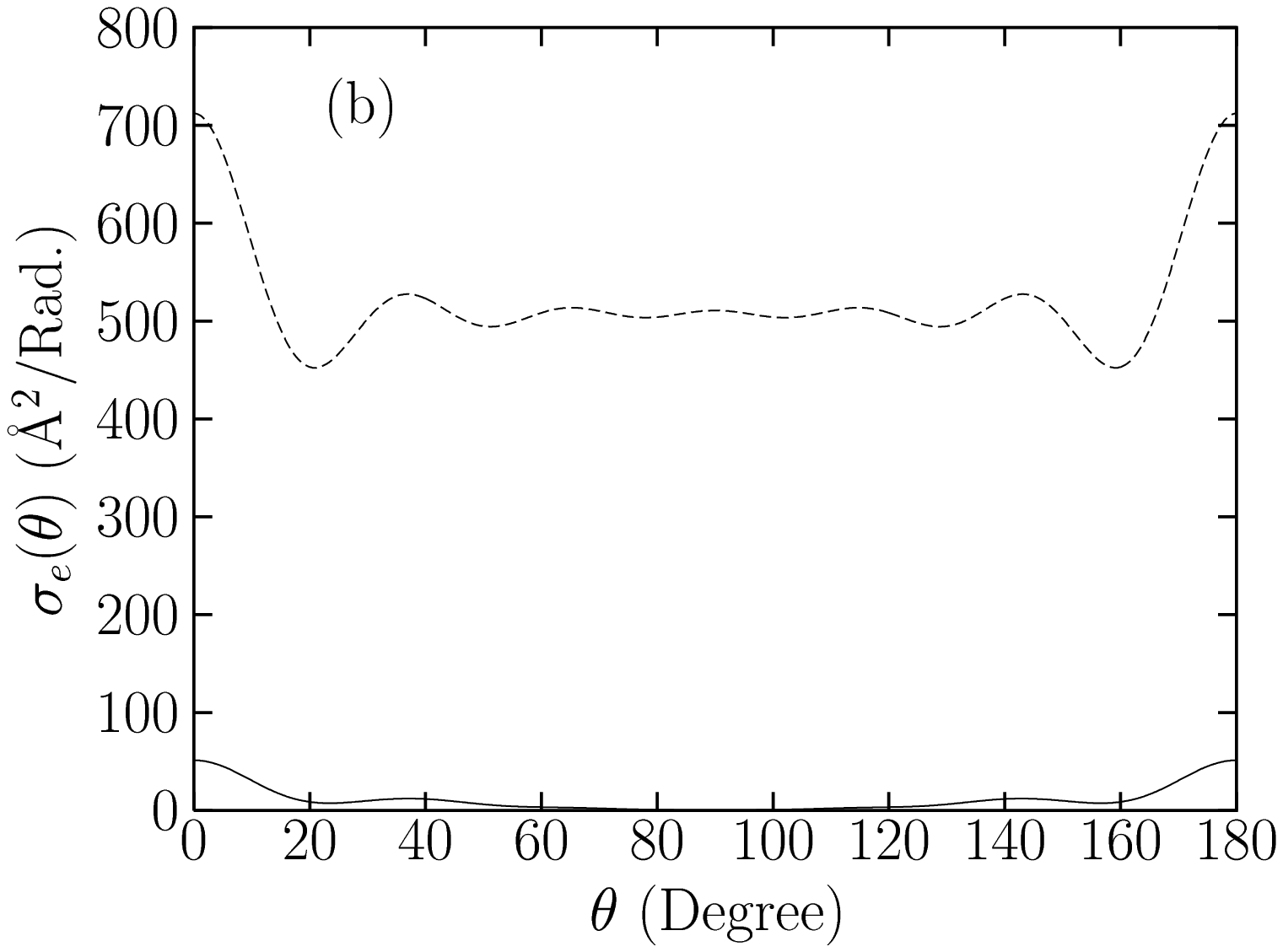,width=6.5cm}
\end{center}
\vspace{-1cm}
\caption{As in  Fig. \ref{Fig5-1} except that elastic collision is ultracold. The
collision energy is (a) $0.4$ cm$^{-1}$,
and (b) $0.04$ cm$^{-1}$.
}
\label{Fig5-3}
\end{figure}

Figure \ref{Fig5-3} shows $\sigma_{e}(\theta)$ for two energies [$0.4$
cm$^{-1}$ and $0.04$ cm$^{-1}$], with the incoming asymptotic states given
by $|\psi_{j_{1}j_{2}}^{\pm}\rangle$ $(j_{1}=2, j_{2}=0)$. In both cases
$\sigma_{e}(\theta)$ for $|\psi_{j_{1}j_{2}}^{+}\rangle$ (dashed line) is
fairly uniform suggesting that s-wave dominates the elastic scattering.
Thus, preventing s-wave scattering via entanglement is expected to cause
a dramatic decrease in the cross section. Indeed, in Fig. \ref{Fig5-3}b,
one sees that elastic scattering is almost totally suppressed for the
$|\psi_{j_{1}j_{2}}^{-}\rangle$ case. Integrating $\sigma_{e}(\theta)$ over $\theta$ gives
$\sigma_{e}^+$ = 724 \AA$^2$ and $\sigma_{e}^-$ = 285 \AA$^2$ with $d_c$ = 87\%
for the case shown in Fig. \ref{Fig5-3}a, and $\sigma_{e}^+$ = 1014 \AA$^2$ and
$\sigma_{e}^-$ = 11 \AA$^2$ with $d_c$ = 195\% for the case shown in Fig.
\ref{Fig5-3}b.

\begin{figure}[ht]
\begin{center}
\epsfig{file=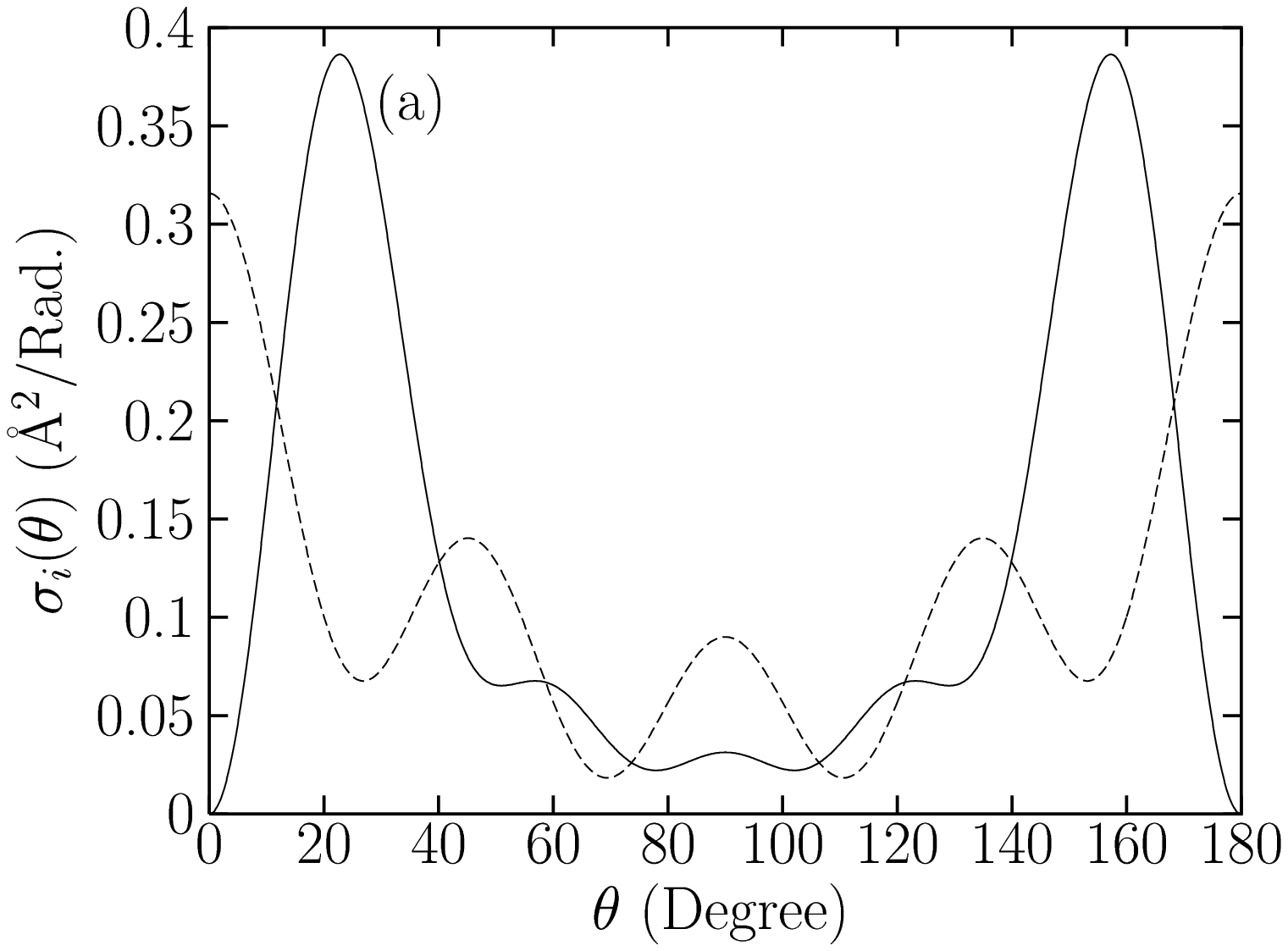,width=6.5cm}

\vspace{-1.4cm}\epsfig{file=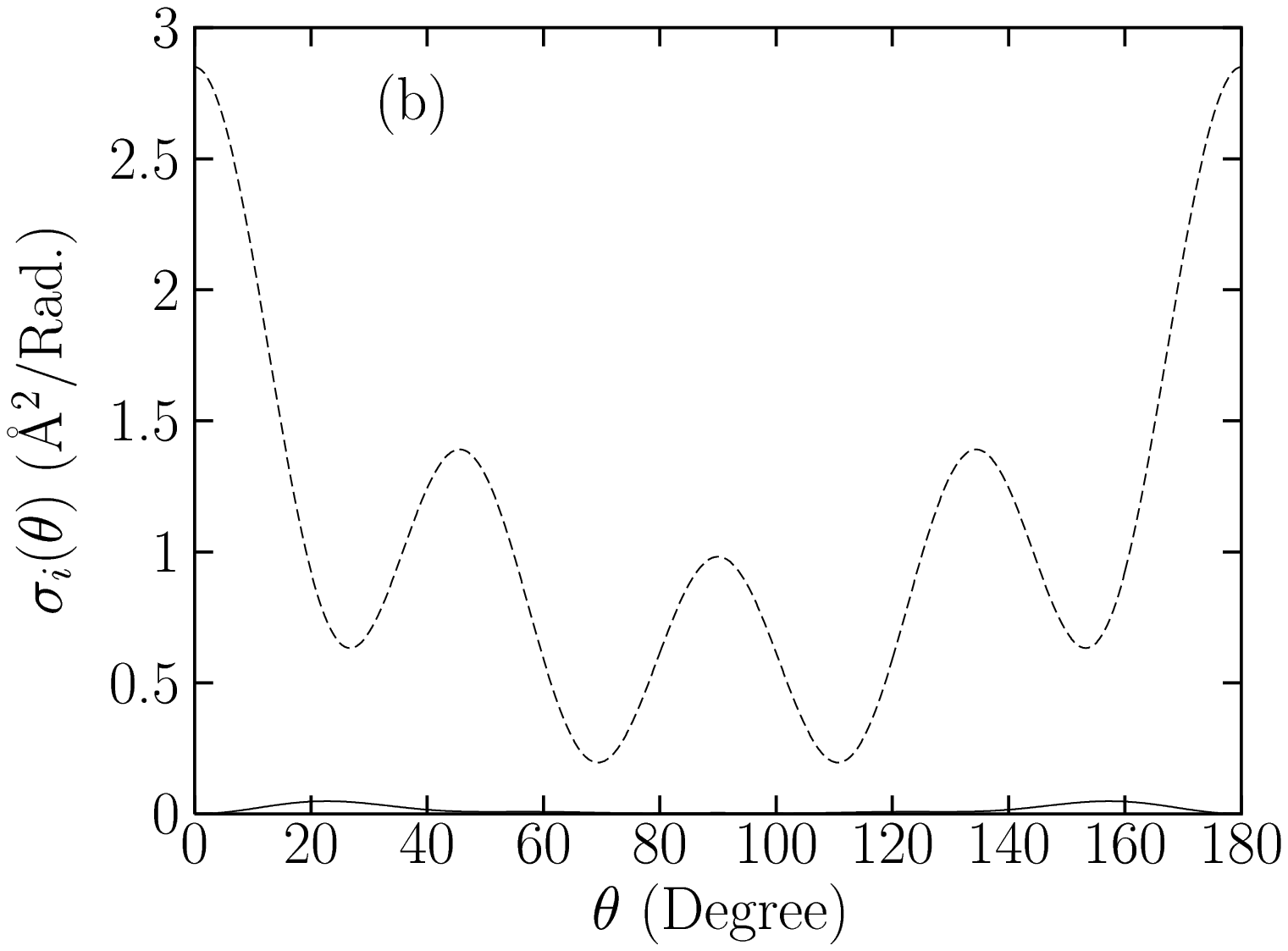,width=6.5cm}
\end{center}
\vspace{-1cm}
\caption{As in Fig. \ref{Fig5-2} except that inelastic collision is ultracold. The
collision energy is (a) $0.04$ cm$^{-1}$,
and (b) $0.0004$ cm$^{-1}$.
}
\label{Fig5-4}
\end{figure}

Figure \ref{Fig5-4}  shows sample results for ultracold inelastic
scattering. As in Fig. \ref{Fig5-2}, $|\psi_{j_{1}j_{2}}^{\pm}\rangle$
$(j_{1}=4, j_{2}=0)$ is chosen as the incoming asymptotic  state and we
examine transitions to $j_{1}'=2, j_{2}'=2$ with collision energies of
$0.04$ cm$^{-1}$ and $0.0004$ cm$^{-1}$. The differences in
$\sigma_{i}(\theta)$ between $|\psi_{j_{1}j_{2}}^{+}\rangle$ and
$|\psi_{j_{1}j_{2}}^{-}\rangle $ in Fig. \ref{Fig5-4}a is somewhat similar
to that in Fig. \ref{Fig5-2}c, where the collision energy is $100$ times
larger, with a corresponding $d_c$ of 22\%. Hence, control is significant,
but the system is insufficiently cold to show {\em extreme} quantum
effects. By contrast,  Fig. \ref{Fig5-4}b shows results for the lowest
$E_{k}$ of $0.0004$ cm$^{-1}$. Clearly, in this case, inelastic scattering
for $|\psi_{j_{1}j_{2}}^{-}\rangle$ is almost completely suppressed,
whereas inelastic scattering for $|\psi_{j_{1}j_{2}}^{+}\rangle$ remains
significant. The integrated cross sections give $\sigma_{i}^+ =1.63 $ \AA$^2$
and $\sigma_{i}^- =0.02 $ \AA$^2$, with a $d_c$=194\%.

Thus, we have demonstrated that quantum entanglement in the incoming
asymptotic state can dramatically alter ultracold collision cross
sections. These effects will also appear in the scattering of the
superposition states in Eq. (\ref{sups}). That is, Eq. (\ref{sigmaeq}) is
well obeyed in all of the cases shown in Figs. \ref{Fig5-1} to
\ref{Fig5-4}. For example, in the elastic scattering case ($j_{1}=2$,
$j_{2}=0$, $v_{1}=v_{2}=0$, $m_{1}=m_{2}=0$, $v_{1}'=v_{2}'=0$,
$j_{1}'=2$, $j_{2}'=0$) the scattering amplitude for the $j_{1}=2,
j_{2}=2$ $\rightarrow j_{1}'=2, j_{2}'=0$ transition due to the first
satellite state is orders of magnitude smaller than that for the $j_{1}=2,
j_{2}=0$ $\rightarrow j_{1}'=2, j_{2}'=0$ transition associated with the
entangled state term at the $E_k$ considered. The probability of the
transition $j_{1}=0, j_{2}=0$ $\rightarrow j_{1}'=2, j_{2}'=0$ due to the
second satellite state is also negligible (indeed, it is exactly zero for
$E_{k}<364.8$ cm$^{-1}$). Similarly, for the inelastic scattering case
($j_{1}=4$, $j_{2}=0$, $v_{1}=v_{2}=0$, $m_{1}=m_{2}=0$,
$v_{1}'=v_{2}'=0$, $j_{1}'=2$, $j_{2}'=2$), the scattering amplitude due
to the first satellite, the de-excitation $j_{1}=4, j_{2}=4$ $\rightarrow
$$j_{1}'=2, j_{2}'=2$, is negligible due
to the large rotational energy mismatch between the initial and final
channels.  In addition, the second satellite state does not contribute to
the product channel $(j_{1}'=2, j_{2}'=2)$ for $E_{k}<729.6$ cm$^{-1}$.

Thus, for example if, in Eq. (\ref{sigmaeq}),
$\alpha_{1}=\alpha_{2}=\pi/4$, then Eq. (\ref{sigmaeq}) reduces to
\begin{eqnarray}
\sigma(\theta,|\psi_{dp}\rangle) & \approx &\frac{1}{2} \sigma(\theta,
|\psi_{j_{1}j_{2}}^{+}\rangle), \ \ \beta=
2n\pi;\nonumber \\
\sigma(\theta,|\psi_{dp}\rangle) & \approx &\frac{1}{2} \sigma(\theta,
|\psi_{j_{1}j_{2}}^{-}\rangle), \ \ \beta= (2n+1)\pi.
\label{b1b2}
\end{eqnarray}
In this case the role of $\beta$ as a control parameter, evident in Eq.
(\ref{b1b2}) as a method for switching between the plus and minus states,
is worthy of note. For example, the elastic scattering results in Fig.
\ref{Fig5-3}b show that by varying $\beta$ one can enhance, or almost
completely suppress, the yield of the $(j_{1}'=2, j_{2}'=0)$ channel in
directions other than that of the incident momentum. Likewise, the
inelastic scattering results in Fig. \ref{Fig5-4}b suggest that by
manipulating $\beta$ we can enhance or almost shut off the $(j_{1}'=2,
j_{2}'=2)$ channel.

\subsection{Vibrational Relaxation}

The above discussion provides results on rotational energy transfer. The
formalism, however, applies equally well to other types of scattering.
Unfortunately, the current state-of-the-computational-art prevents a full
quantum calculation on reactive scattering, or even on diatom-diatom
rovibrational energy transfer. That is, the exact numerical treatment of
$AB$ + $AB$ in three dimensions is still not feasible and approximations
such as the sudden approximation \cite{clary} may not be appropriate for
low energy collisions of the type that we are examining since it assumes
that molecular rotation is slow compared with that of vibration and
translation. Further, although a semiclassical treatment of rovibrational
energy transfer is available \cite{billing} it ignores the role of the
quantum effects in the translational motion. Nonetheless, we wish to gain some
insight into phase control of vibrational relaxation. For this reason we
present results on a simplified model of diatom-diatom scattering where
the rotational motion is frozen.
Specifically, we adopt a three-degree-of-freedom model that assumes that
both the projectile and target molecules point at a fixed direction during
the entire scattering process. The scattering problem is then
computationally solved by wavepacket propagation.  To do so we employ a
time-dependent approach with real $L^{2}$ eigenfunctions with damping
\cite{bowman,reall2note}.

\begin{figure}[ht]
\begin{center}
\ \

\epsfig{file=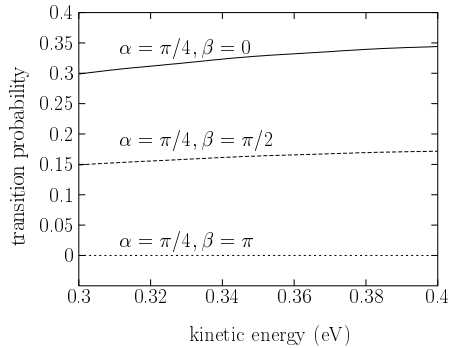,width=6.5cm}
\end{center}
\vspace{-1cm}
\caption{The inelastic transition probability of the channel $(v_{1}'=1,
v_{2}'=1)$ versus the collision energy in the para-  H$_{2}
$ + para- \ H$_2$ scattering,
for three incoming entangled vibrational states described by
Eq. (\ref{ven}). $\alpha=\pi/4$, $\beta=0$ (top curve), $\pi/2$ (middle
curve) and  $\pi$ (bottom curve).  The internuclear axis of each
para--H$_2$ molecule is assumed to be parallel to the incident velocity
during the scattering process.
}
\label{Fig5-5}
\end{figure}

Figure \ref{Fig5-5} displays the transition probability to the product
$(v_{1}'=1, v_{2}'=1)$ channel for three initial entangled vibrational
states characterized by
\begin{eqnarray}
\psi_{v_1v_2}^{(\alpha,\beta)} = \cos(\alpha)|v_{1}=0\rangle\otimes
|v_{2}=2\rangle + \exp(i\beta) \sin(\alpha)|v_1=2\rangle\otimes
|v_{2}=0\rangle
\label{ven}
\end{eqnarray}
with $\alpha=\pi/4$, $\beta=0,\ \pi/2,$ and $\pi$, respectively. As seen
in Fig. \ref{Fig5-5}, the  quantum phase $\beta$ embedded in the initial
entangled vibrational states strongly affects vibration-vibration
relaxation, with the cross section changing considerably as a function of
the initial entangled state.

\section{Discussion and Summary}
\label{s5-4}

The results on para--H$_2$ + para--H$_2$ scattering make clear that
entanglement can play a significant role in the control of nonreactive
collisional processes. To investigate this phenomenon experimentally one
could attempt to prepare the entangled molecular state directly. However,
although quantum entanglement of atomic systems has been experimentally
realized \cite{entangleexp,enbeams}, preparing entangled molecular states
would require a considerable extension of technology. By contrast,
contributions from entangled states appear naturally in the collision of
AB + AB systems where the initial states are prepared as superpositions of
the form given in Eq. (\ref{sups}). Preparing these states would entail
excitation of the $|j_1,m_1,v_1 \rangle$ to produce the superposition with
$|j_2,m_2,v_2\rangle$. If the transition is dipole allowed then direct
excitation of the lower state is possible. Alternatively, stimulated Raman
adiabatic passage (STIRAP) \cite{bergmann1} or the tripod-STIRAP scheme
\cite{bergmann2} may provide a useful choice for preparing the
superposition states for both the projectile and the target molecules. In
the latter approach an additional laser couples the intermediate level
(through which the initial and final state are radiatively connected via
the pump and Stokes-laser) with another unpopulated state.  Depending on
the time overlap of the additional laser with that of the pump and
Stokes-laser, any coherent superposition state of the initial and final
state can be created.  Thus, by introducing different coherence
characteristics from laser fields into the scattering system, we can
control $\beta$ and thus select the form of the entangled state component
$|\psi_{dp}\rangle$ in Eq. (\ref{psidp}), giving rise to phase control
over differential and total cross sections.

Our studies have been restricted to nonreactive scattering of AB + AB,
with particular numerical application to para--H$_2$ + para--H$_2$. As
such, it remains to discuss the potential for applications to inelastic
scattering of other systems and to comment on control of reactive
scattering of identical particles.

{\it General inelastic $AB$ + $AB$ scattering:} Extensions to other
inelastic scattering cases fall into two categories. The first are systems
which, like para--H$_2$ + para--H$_2$,  have zero total nuclear spin,
zero total electronic orbital angular momentum, and zero total electronic spin.
The theory of these cases is described above and is directly applicable to
inelastic scattering of other zero spin cases such as $CO$ + $CO$, a
system whose vibrational energy transfer is of interest in laser physics.
However, the choice of zero spin merely simplifies theoretical
considerations and it is not essential to control.

The second category, molecules with nonzero nuclear spin or electronic
spin, does require an extension of the theory provided above. For such
molecules, one of the two quantum effects due to entangled molecular
states, i.e., the parity selection of the incoming partial waves, may be
substantially reduced when the nuclear or electronic spin is unpolarized.
To see this note that the spin degree of freedom can introduce additional
permutation symmetries into the system. Thus, entangled rovibrational
states like $|\psi(j_{1}v_{1}m_{1}j_{2}m_{2}v_{2})\rangle_{\pm}$ do not
necessarily select the partial waves. For instance,  for two unpolarized
identical molecules with total integer nuclear spin $I_{N}$, the
permutation symmetry of all other degrees of freedom is $+1$ with the
probability $(I_{N}+1)/(2I_{N}+1)$ and $-1$ with the probability
$(I_{N})/(2I_{N}+1)$ \cite{kouri}. When $I_{N}$ is large, two permutation
symmetries of all other degrees of freedom are equally allowed.  As a
result, the permutation symmetry induced by entanglement of molecular
rovibrational motion will have a negligible effect on the differential or
total cross sections. However, as suggested  in Ref. \cite{pauljcp}, one
quantum effect still survives, i.e., quantum interference between the
transition amplitudes
$T^{JM}(j_{1}'v_{1}'j_{2}'v_{2}'j_{12}'l'|j_{1}v_{1}j_{2}v_{2}j_{12}l)$
and
$T^{JM}(j_{1}'v_{1}'j_{2}'v_{2}'j_{12}'l'|j_{2}v_{2}j_{1}v_{1}j_{12}l)$.
Establishing the magnitude of this effect alone will require further
study.

Finally, note that there are considerable differences between the
inelastic scattering case studied here and the case of reactive AB + AB
scattering (to form A$_2$B + B or A$_2$ + B$_2$), the subject of future
research.  A full treatment of reactive scattering, which does require
considerable extensions of current computational technique, is under
consideration.

In summary, we have examined the nature of the interference, and hence
control, in nonreactive AB + AB scattering when the
initial states are in a quantum superposition state. In doing so we have
exposed the relationship of the interference term to entangled molecular
rovibrational states as the incoming asymptotic state. Intriguing quantum
effects resulting from quantum entanglement are revealed, and are shown to
provide a novel means of controlling both nonreactive 
differential and total cross
sections in identical diatom-diatom scattering.

\vspace{0.5cm}
{\bf Acknowledgments}: 
The authors thank Dr. S. Skokov and Prof. J. Bowman for
providing their scattering code (used for the vibrational relaxation case)
employing the time-dependent approach with real $L^{2}$ eigenfunctions
with damping. This work was supported by the U.S. Office of Naval Research
and the Natural Sciences and Engineering Research Council of Canada.

\pagebreak


\begin{thebibliography}{100}
\bibitem{paulreview} M. Shapiro and P. Brumer, Adv. Atom. Mol. and Opt. Phys. {\bf 42},
287 (2000); M. Shapiro and P. Brumer, {\it Principles of the Quantum Control
of Molecular Processes} (John Wiley, New York, in press).
\bibitem{rice}  S.A. Rice and M. Zhao, {\it Optical Control of Molecular Dynamics} (John
Wiley,  New York, 2000).
\bibitem{rabitz} 
H. Rabitz, R. de Vivie-Riedle, M. Motzkus, and K. Kompa, Science {\bf 288},
824 (2000).
\bibitem{sanchez} V. Sanchez-Villicana., S.D. Gensemer, K.Y.N. Tan, A. Kumarakrishnan,
T.P. Dinneen, W. Suptitz,
    and P.L. Gould, \prl{\bf 74}, 4619 (1995).
\bibitem{fedichev} P.O. Fedichev, Yu. Kagan, G.V. Shlyapnikov,  and J.T.M.
Walraven, \prl{\bf 77}, 2913 (1996).
\bibitem{dcfield}M. Marinescu and L. You, \prl{\bf 81}, 4596 (1998).
\bibitem{paulsca} M. Shapiro and P. Brumer, \prl{\bf 77}, 2574 (1996).
\bibitem{lasercat} 
M. Shapiro and Y. Zeiri, J. Chem. Phys. {\bf 85}, 6449 (1986); 
T. Seideman and M. Shapiro, J. Chem. Phys., {\bf 94}, 7910 (1991); 
A. Vardi and M. Shapiro, Comm. At. Mol. Phys. {\bf 2}, D233 (2001).
\bibitem{alex} A. Abrashkevich, M. Shapiro, and P. Brumer, \prl{\bf 81},
3789 (1998); Erratum: \prl{\bf 82}, 3002 (1999); Faraday Discuss. {\bf 113}, 291 (1999);
E. Frishman, M. Shapiro, and P. Brumer, J. Phys. Chem. A {\bf 103}, 10333
(1999);
A. Abrashkevich, M. Shapiro, and P. Brumer, Chem. Phys. {\bf 267}, 81 (2001).
\bibitem{pauljcp}P. Brumer, K. Bergmann, and M. Shapiro, J. Chem. Phys. {\bf 113}, 2053
(2000).
\bibitem{qcom}
M.A. Nielsen and I.L. Chuang, {\it Quantum Computation and Quantum
Information} (Cambridge University Press, Cambridge, 2000).
\bibitem{qtele} C.H. Bennett, G. Basare,
C. Cr\'{e}peau,
R. Jozsa, A. Peres,
and W.K. Wootters, \prl{\bf 70}, 1895 (1993).
\bibitem{kurizki} For a recent teleportation scenario based on entanglement in
molecular dissociation and collisions, see T. Opatrny and G.
Kurizki, \prl{\bf 86}, 3180 (2000).
\bibitem{ghirardi} G. Ghirardi, L. Marinatto, and T. Weber, J. Stat. Phys. {\bf
108}, 49 (2002).
\bibitem{taylor51} J.R. Taylor, {\it Scattering Theory} (Wiley, New York, 1972).
\bibitem{green}S. Green, J. Chem. Phys. {\bf 62}, 2271 (1975).
\bibitem{kouri}T.G. Heil,  S. Green, and D.J. Kouri, J. Chem. Phys. {\bf 68}, 2562 (1978).
\bibitem{polartheory} B.A. Robson, {\it The Theory of Polarization Phenomena} (Clarendon
Press, Oxford, 1974).
\bibitem{parity-note}To prove the equivalence between Eq. (\ref{ftmatrix2})
and
Eq. (\ref{ftmatrix}), one needs to use the fact that
due to parity conservation,
the $T$-matrix elements
$T^{JM}(j_{1}'v_{1}'j_{2}'v_{2}'j_{12}'l'|j_{1}v_{1}j_{2}v_{2}j_{12}l)$
are nonzero only if
$(-1)^{j_{1}+j_{2}+l}=(-1)^{j_{1}'+j_{2}'+l'}$.
\bibitem{molscat} J. M. Hutson and S. Green, MOLSCAT computer code, version 14 (1994),
distributed by  Collaborative Computational Project No. 6 of the Engineering and Physical
Research Council (UK).
\bibitem{Rabitz2} G. Zarur and H. Rabitz, J. Chem. Phys. {\bf 60}, 2057
(1974).
\bibitem{4hpotential} A. Aguado, C. Su\'{a}rez, and M. Paniagua, J. Chem. Phys. {\bf 101},
4004 (1994).
\bibitem{vlado} V. Zeman, University of Toronto (private communication).

 \bibitem{dalgarno} N. Balakrishnan, R.C. Forrey, and A. Dalgarno, \prl{\bf 80}, 3224
 (1998);
 R.C. Forrey, N. Balakrishnan, A. Dalgarno,
 M.R. Haggerty, and E.J. Heller,  \prl{\bf 82}, 2657 (1999);
 E. Bodo, F.A. Gianturco, and A. Dalgarno, J. Chem. Phys. {\bf 116}, 9222 (2002).
 \bibitem{superchem} D.J. Heinzen, R. Wynar, P.D. Drummond, and K.V. Kheruntsyan,
 \prl{\bf 84}, 5029 (2000).
 \bibitem{bohn} A.V. Avdeenkov and J.L. Bohn, \pra{\bf 64}, 052703 (2001); J.L. Bohn, \pra{\bf 62},
 032701 (2000); A. Volpi and J.L. Bohn, \pra{\bf 65}, 052712 (2002).
 \bibitem{zare} R.N. Zare first suggested applying
 phase control to ultracold collisions in  Faraday Discuss. Chem.
 Soc. {\bf 113}, p351 (1999).
\bibitem{note2} Even in the zero collision energy limit,
it is still possible that many partial waves contribute to the
cross section, e.g., when the interaction potential approaches
zero as $1/R^{3}$ \cite{dcfield}.
\bibitem{note3}
In the close-coupling calculations, it is found that, even though the total
cross section has quickly converged,  substantially increasing the cut-off value of the
total angular momentum may cause some artificial oscillations
in the differential cross sections for $\theta$ very close to $0$ or $\pi$.
\bibitem{clary} D.C. Clary, Chem. Phys. Lett. {\bf 74}, 454 (1980);
R. Hern\'{a}ndez, R. Toumi, and D. Clary, J. Chem. Phys. {\bf 102}, 9544
(1995).
\bibitem{billing} V.A. Zenevich and G. Billing, J. Chem. Phys. {\bf 111}, 2401 (1999).
\bibitem{bowman}  S. Skokov and J. M. Bowman,  Phys. Chem. Chem. Phys. {\bf 2}, 495
(2000).
\bibitem{reall2note} Direct wavepacket propagation is a powerful tool to
extract scattering information.  Wavepacket propagation would be formally
and easily done if the scattering eigenfunction and eigenenergy in the
continuum are known. The central idea of the approach in Ref.
\cite{bowman} is to  replace the non-$L^{2}$ scattering eigenstates by
real $L^{2}$ states which can be numerically treated.  To do this, the
continuum is discretized in a box and real eigenstates are obtained.  The
wavepacket evolution can then be carried out almost trivially.   To avoid
unphysical reflections at the grid edges, the wavepacket is damped at each
time step.  In obtaining eigenfunctions, a potential-optimized discrete
variable representation is employed,  and the truncation/recoupling
technique is essential for efficient calculations.
\bibitem{entangleexp} E. Hagley, X. Ma\^{i}tre, G. Nogues, C. Wunderlich, M. Brune,
J.M. Raimond, and S.
Haroche, \prl{\bf 79}, 1 (1997); Q.A. Turchette,
C.S. Wood, B.E. King, C.J. Myatt, D. Leibfried, W.M. Itano, C. Monroe, and D.J.
    Wineland,
     \prl{\bf 81}, 3631 (1998); B. Julsgaard, A. Kozhekin, and E.S. Polzik, Nature
     {\bf
      413}, 400 (2001).
\bibitem{enbeams} H. Pu and  P. Meystre, \prl{\bf  85}, 3987 (2000);
      L.-M. Duan, A. S$\phi$rensen, J.I. Cirac, and P. Zoller, \prl{\bf  85}, 3991
      (2000).
\bibitem{bergmann1} K. Bergmann, H. Theuer, and B.W. Shore, Rev. Mod. Phys.
      {\bf 70}, 1003
      (1998).
\bibitem{bergmann2} R.G. Unanyan, M. Fleischhauer, K. Bergmann, and B.W.
      Shore, Opt.
      Commun. {\bf 155}, 144 (1998); H. Theuer, R.G. Unanyan, C. Habscheid, K.
      Klein, and
      K. Bergmann, Optics Express {\bf 4}, 77 (1999).

\end{thebibliography}
\end{document}